\pdfoutput=1
\documentclass[reprint,amsmath,amssymb,aps,floatfix]{revtex4-1}

\usepackage[utf8]{inputenc}
\usepackage[english]{babel}
\usepackage[export]{adjustbox}
\usepackage[svgnames]{xcolor}
\usepackage{wrapfig}  % Another package for wrapping text - just for figures
\usepackage{tabularx}
\usepackage{xr-hyper}
\usepackage{hyperref} 
\usepackage{bm}
\usepackage{url} 
\usepackage[normalem]{ulem}
\usepackage{CJK}
\usepackage{multirow}

\usepackage[abs]{overpic}
\usepackage{mdframed}

\newcommand{\Gammap}[0]{\mathit{\Gamma}} % TODO italics
\newcommand{\Kp}[0]{\mathit{K}}
\newcommand{\Mp}[0]{\mathit{M}}
\newcommand{\Xp}[0]{\mathit{X}}

\newcommand{\TheSinglet}[0]{$A'_1$}
\newcommand{\vect}[1]{\bm{#1}}
\newcommand{\abs}[1]{\left|{#1}\right|}
\newcommand{\ush}[2]{Y_{#1,#2}}
\newcommand{\svwfr}[3]{\mathbf{u}_{#1,#2}^{#3}}
\newcommand{\svwfs}[3]{\mathbf{v}_{#1,#2}^{#3}}
\newcommand{\coeffr}[0]{p}
\newcommand{\coeffs}[0]{a}

\newcommand{\coeffsip}[4]{\coeffs_{#1}^{#2,#3,#4}}

\newcommand{\coeffrip}[4]{\coeffr_{#1}^{#2,#3,#4}}
\newcommand{\coeffripext}[4]{\coeffr_{\mathrm{ext}(#1)}^{#2,#3,#4}}
\newcommand{\transop}[0]{S}

\newcommand{\PT}[1]{\textcolor{Green}{{\bf PT: #1 }}}

\newcommand{\MMNmark}[1]{\textcolor{red}{MMN: #1}}

\newcommand{\diffnote}[1]{} 	% do not print in the final version
\newcommand{\changed}[1]{#1}  % just print in the final version
\newcommand{\removed}[1]{}	    % do not print in the final version
\definecolor{BLUE}{cmyk}{0.9998,1,0,0} % a hack to make the colour work also in section headings

\newcommand{\altcombinedimage}[2]{#2}	% standard mode, use subfigs directly
\newcommand{\arxivonly}[1]{{#1}}	% For arXiv etc.
\renewcommand{\PT}[1]{\changed{#1}}

\renewcommand{\MMNmark}[1]{\changed{#1}}

\renewcommand{\PT}[1]{} % remove the notes

\begin{document}
\author{R.\ Guo$^1$}
\author{M.\ Nečada$^1$}
\author{T.K.\ Hakala$^{1,2}$}
\author{A.I.\ Väkeväinen$^1$}
\author{P.\ Törmä$^1$}
\email{paivi.torma@aalto.fi}
\affiliation{$^1$Department of Applied Physics, Aalto
University, FI-00076 Aalto, Finland \\
$^2$Institute of Photonics, University of Eastern Finland, P.O.~Box 111, FI-80101 Joensuu, Finland}

\title{Lasing at the $\Kp$-points of a honeycomb plasmonic lattice}

\begin{abstract}
We study lasing at the high-symmetry points of the Brillouin zone in a
honeycomb plasmonic lattice. We use symmetry arguments to define
singlet and doublet modes at the $\Kp$-points of the
reciprocal space. We experimentally demonstrate lasing at
the $\Kp$-points \changed{that is based on plasmonic lattice modes and
two-dimensional feedback}. By comparing polarization properties to
\changed{$T$-matrix} simulations, we identify the lasing mode as one of
the singlets with an energy minimum at the $\Kp$-point enabling
feedback.  Our results offer prospects for studies of topological
lasing in radiatively coupled systems. 

\end{abstract}

\maketitle

Feedback provided by a resonator is essential for lasing. 
The resonator can be a set of mirrors~\cite{Koyama1987} {or}
periodic structures enabling distributed feedback (DFB)
lasing~\mbox{\cite{Kogelnik1972,Dowling1994,Meier1998,Noda2001,Matsubara2008}}. Most
DFB lasers rely on simple one-dimensional periodic structures. More complex
geometries would offer such interesting features as distributed feedback
involving multiple modes, flat bands, and increased variety of degenerate
high-symmetry points and possibilities of creating topological
bands~\cite{ozawa_topological_2018}.  The symmetry of a hexagonal Bravais
lattice leads to the possibility to multiply degenerate points at the first
Brillouin zone edge {\cite{dresselhaus_group_2008}}. Here we experimentally
demonstrate lasing at $\Kp$-points of a honeycomb plasmonic lattice. 

The vast majority of the work on bosons in hexagonal/honeycomb lattices, for
photonic \cite{Weick2013,Rechtsman2013,Jacqmin2014}, microwave
\cite{Bittner2010,Bellec2013}, and atomic
\cite{Becker2010,Chen2011,Struck2013,Li2016} systems realize essentially the
tight-binding model of the lattice. That is, the lattice sites are connected
only up to the (next-)nearest neighbor; in the optical systems, this is
realized by site-to-site near-field coupling. Our system consists of an array
of plasmonic nanoparticles that are radiatively coupled over the whole system
size. This renders tight-binding models useless, and we base our theoretical
description on symmetry arguments and $T$-matrix scattering simulations. 

Plasmonic nanohole and nanoparticle arrays combined with organic and inorganic
gain materials are emerging as a versatile platform for room-temperature,
ultrafast
lasing~\cite{Zhou2013,VanBeijnum2013,Meng2014,schokker_lasing_2014,Yang2015,cuerda_theory_2015,
Hakala2017,Wang2017a,ramezani_plasmon-exciton-polariton_2017,schokker_systematic_2017,Rekola2018,daskalakis_ultrafast_2018,tenner_two-mode_2018,Wu2016,Zhang2015} and Bose-Einstein condensation
\cite{Martikainen2014,hakala_bec_2018}. These works, however, focus on
lasing action or condensation at the $\Gammap$-point, that is, at the center of
the Brillouin zone of systems with a Bravais lattice that is
rectangular/square~\cite{Zhou2013,VanBeijnum2013,Meng2014,schokker_lasing_2014,Yang2015,Hakala2017,Wang2017a,ramezani_plasmon-exciton-polariton_2017,schokker_systematic_2017,daskalakis_ultrafast_2018},
hexagonal~\cite{tenner_two-mode_2018,Zhang2015} or
one-dimensional~\cite{Rekola2018} (ref.~\cite{Wu2016} studies lasing action in
the $\Xp$-point of a square lattice). 

$\Kp$-point lasing or condensation in radiatively (long-range) coupled
hexagonal/triangular lattices has been studied in photonic
crystal~\cite{Notomi2001, Chen2010, Huang2016} and
exciton-polariton~\cite{Kim2013} systems. In those works, however, the
polarization properties of the output light were not analyzed. Here we
demonstrate lasing at the $\Kp$-points and show
that the polarization properties and real-space patterns of the laser emission
contain essential information about the lasing mode. {We identify the lasing
mode as one of the singlets allowed by symmetry and explain why this mode is
selected by the lasing action.} 

\begin{figure}[h] \centering
	\includegraphics[width=1\columnwidth]{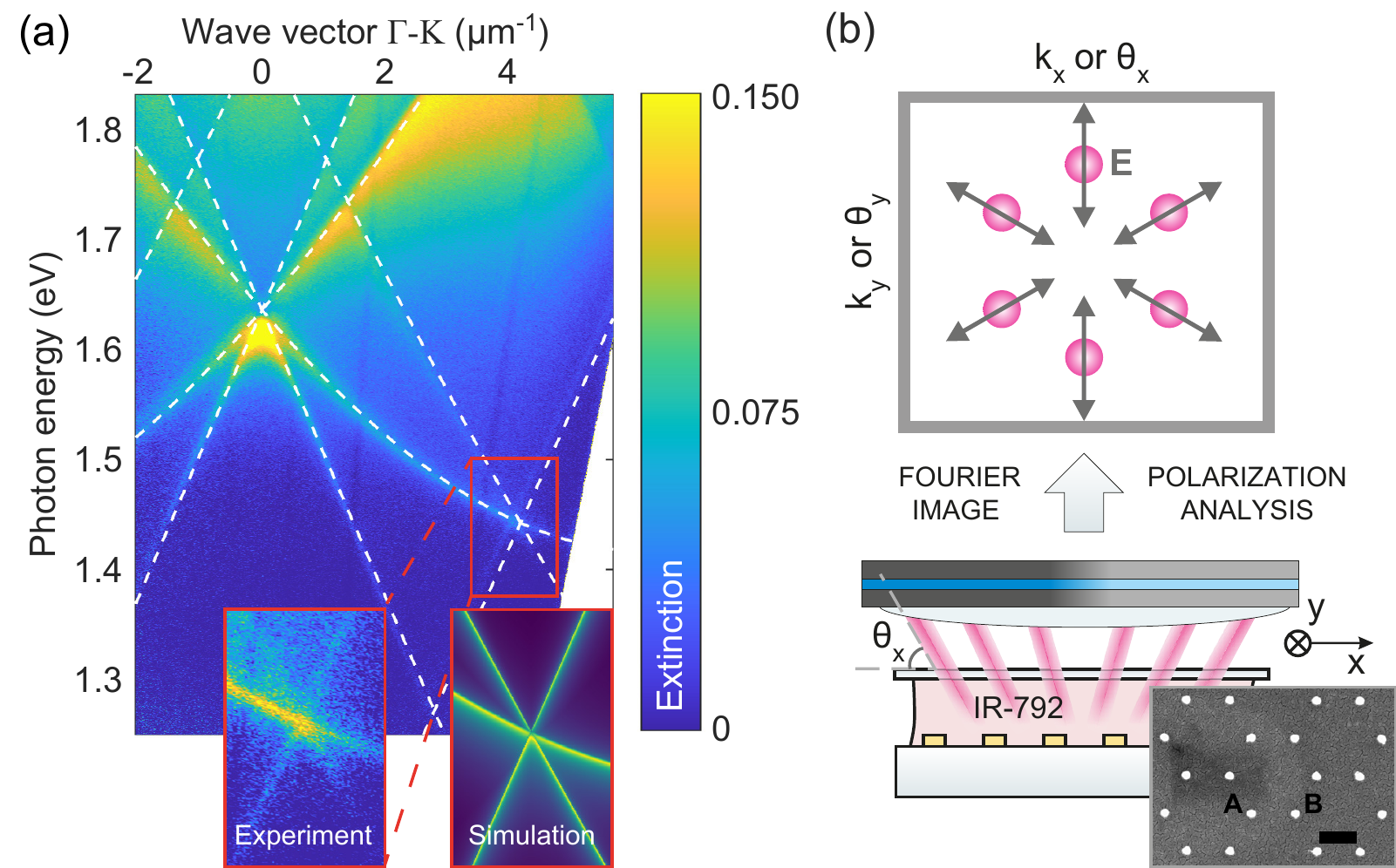}
	\caption{	{(a)} A measured angle-resolved extinction spectrum of a honeycomb
	lattice with particle separation of $p=576\,\mathrm{nm}$. \changed{Color scale shows 
	the extinction which is defined as 
	$(1 - \mbox{normalized transmission})$.} {The SLR
	modes correspond to extinction maxima, closely following the diffracted
	orders (dashed lines).} The left inset shows the measured dispersion around the $\Kp$-point (the
	color scale is from 0 to 0.05). The right inset shows the dispersion
	obtained by $T$-matrix simulations.  {(b) The lasing measurements.
	Nanoparticle samples combined with IR-792 molecules in solution are
	pumped with a femtosecond laser. The hexagonal geometry of the lattice
	(inset: \removed{SEM}\changed{scanning electron microscope}
	image of the gold nanoparticles, scale bar 500~nm, with the
	A and B unit cell sites marked) enables lasing emission in six distinct
	off-normal angles, collected by a 0.6 NA objective and further
	analyzed. In the Fourier image, the six angles correspond to lasing at
	the six $\Kp$-points of the first Brillouin zone, with distinct
	polarization directions (grey arrows) of the electric field
	$\vect{E}$.} } \label{Fig1} \end{figure}

We fabricate cylindrical gold nanoparticles with electron-beam
lithography on a glass substrate {in a honeycomb lattice arrangement}.
The particle
separation is varied between $569–583\,\mathrm{nm}$.  Individual nanoparticles
have a nominal diameter of $100\,\mathrm{nm}$ and height of  $50\,\mathrm{nm}$.
An organic dye molecule IR-792 is added on top of the array in 25~mM solution
and the structure is sealed with a glass superstrate {(Fig.~\ref{Fig1}(b))}.
The dye molecules act as the gain material and are optically pumped
with 100~fs laser pulses (750~nm central wavelength). For  details, see
\arxivonly{section \ref{sm:methods} of }Supplemental Material. 

The energies of diffracted orders (DOs) of a 2D hexagonal lattice are shown
with dashed lines in Fig.~\ref{Fig1}(a) for the $\Gammap$–$\Kp$ in-plane ($x$–$y$
plane) momentum direction. The DOs correspond to diffraction without resonant
phenomena at the lattice sites, so-called empty lattice approximation. In our
samples, the nanoparticles have a broad plasmonic resonance (at 1.87 eV, width $\sim300$
meV) which hybridizes with the DOs, leading to narrow
(width 5–20 meV) dispersive modes called surface lattice resonances
(SLRs)~\cite{Zou2005a,Wang2018}, Fig.~\ref{Fig1}(a). A dispersion obtained by
multiple-scattering $T$-matrix simulation (for details, see \cite{Hakala2017} and
\arxivonly{section \ref{sm:tmatrix} of }supplemental Material) agrees with the experiments, see the insets of
Fig.~\ref{Fig1}(a). The dispersions are measured with a Fourier imaging setup
used in our previous works~\cite{Hakala2017,Guo2017,hakala_bec_2018} but now
extended to larger angles. 

The geometry of an infinite honeycomb lattice belongs to the group $p6m \times
\sigma_h$, the wallpaper group $p6m$ \cite{ITfC:A} extended by the horizontal
reflection $\sigma_h$.  The horizontal reflection ensures that the eigenmodes
can be divided into two classes according to the electric field \changed{orientation} at the mirror
plane: the electric field $\vect{E}$ is either parallel (in-plane-$\vect{E}$, the magnetic
field $\vect{H}$ is then perpendicular to the mirror plane) or perpendicular
(perpendicular-$\vect{E}$, magnetic field $\vect{H}$ in-plane)~\cite{dresselhaus_group_2008}.

\begin{figure} \begin{center}
\altcombinedimage{eigenmodes-theory.pdf}{\begin{tabular}{ccc}  (a) & (b) & \multirow{2}{*}{
		\begin{minipage}[c]{.38\columnwidth}
		\noindent
		\begin{overpic}[width=1\columnwidth]{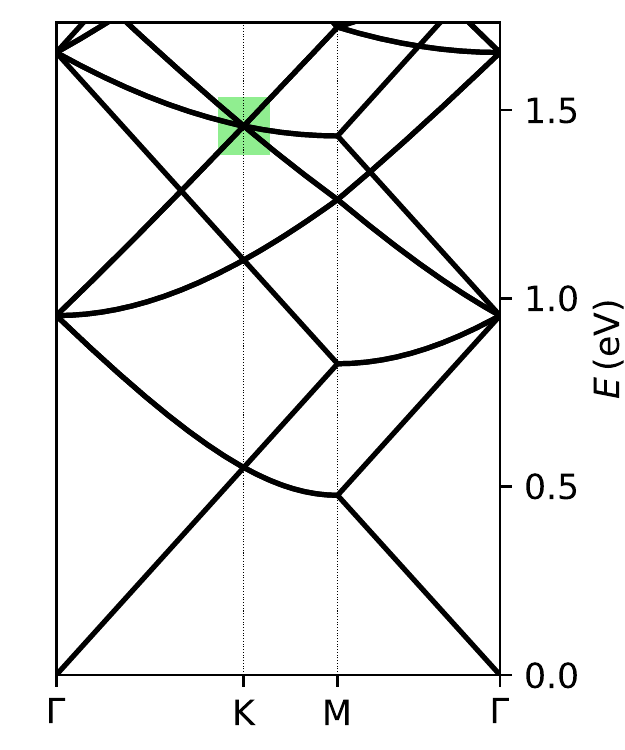}
			\put (8,101) {\setlength{\fboxsep}{1pt}{\colorbox{blue!20}{(c)}}}
		\end{overpic}\\	
		\begin{overpic}[width=1\columnwidth]{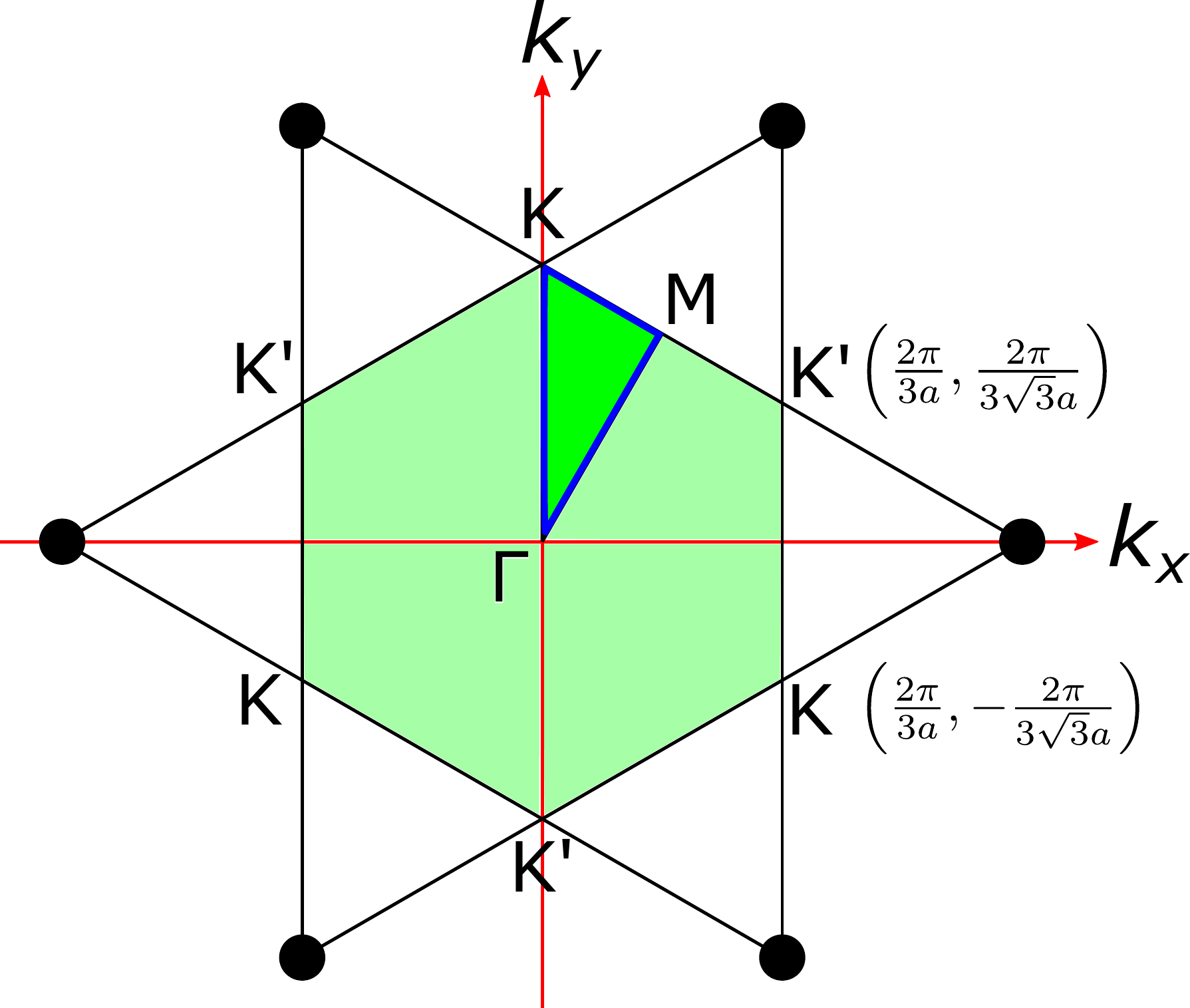}
			\put (3,63) {(d)}
		\end{overpic}\\
	    % generated with qpms/ipynotebooks/hexlaser/Dispersions-projections_v2-fig2c-crosscut.ipynb
		\begin{overpic}[width=1\columnwidth]{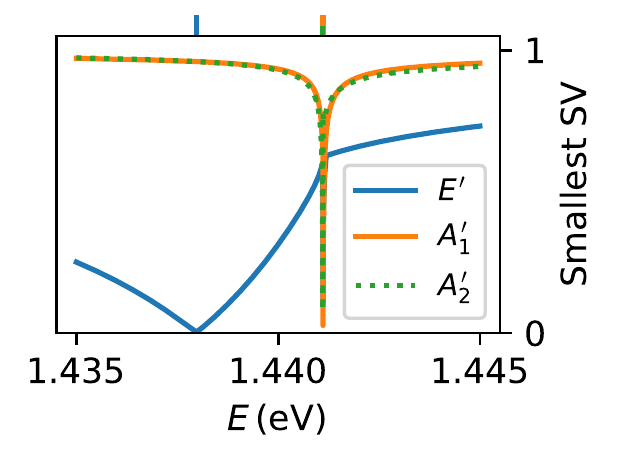}
			\put (10,48) {(e)}
		\end{overpic}\\
	\end{minipage}
		}\\%
	\begin{minipage}[c]{.24\columnwidth}
		\includegraphics[width=1\columnwidth]{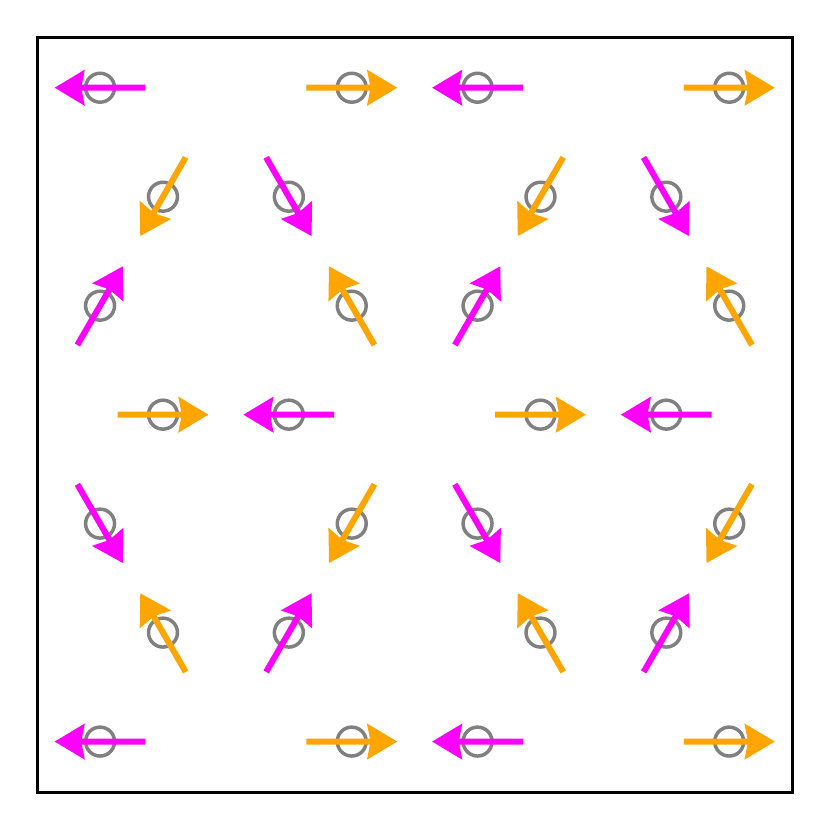}\\
		\includegraphics[width=1\columnwidth]{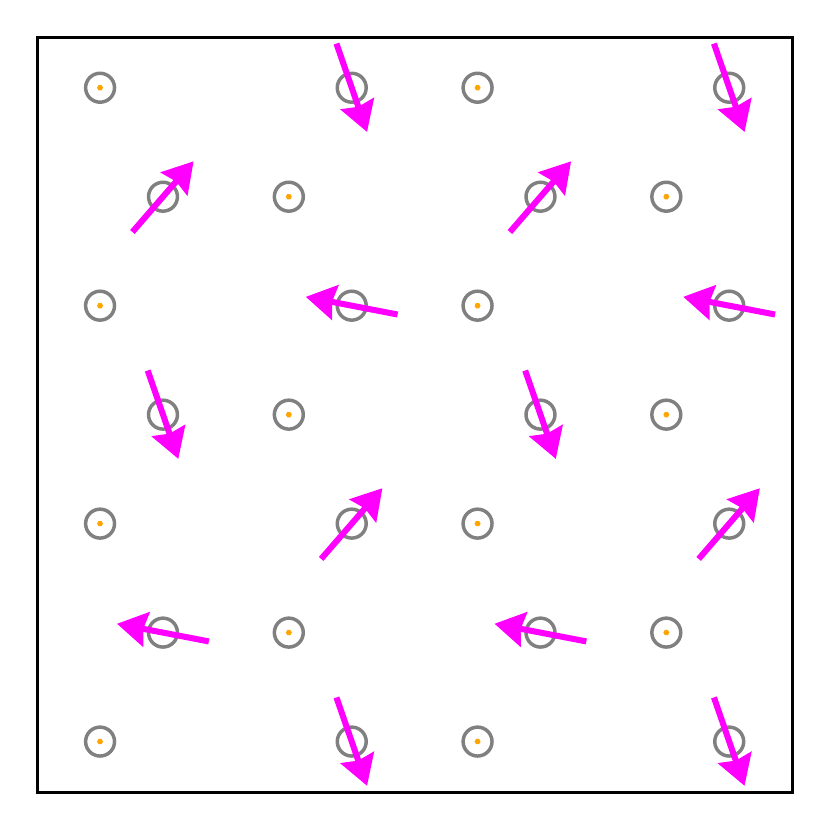}\\
		\includegraphics[width=1\columnwidth]{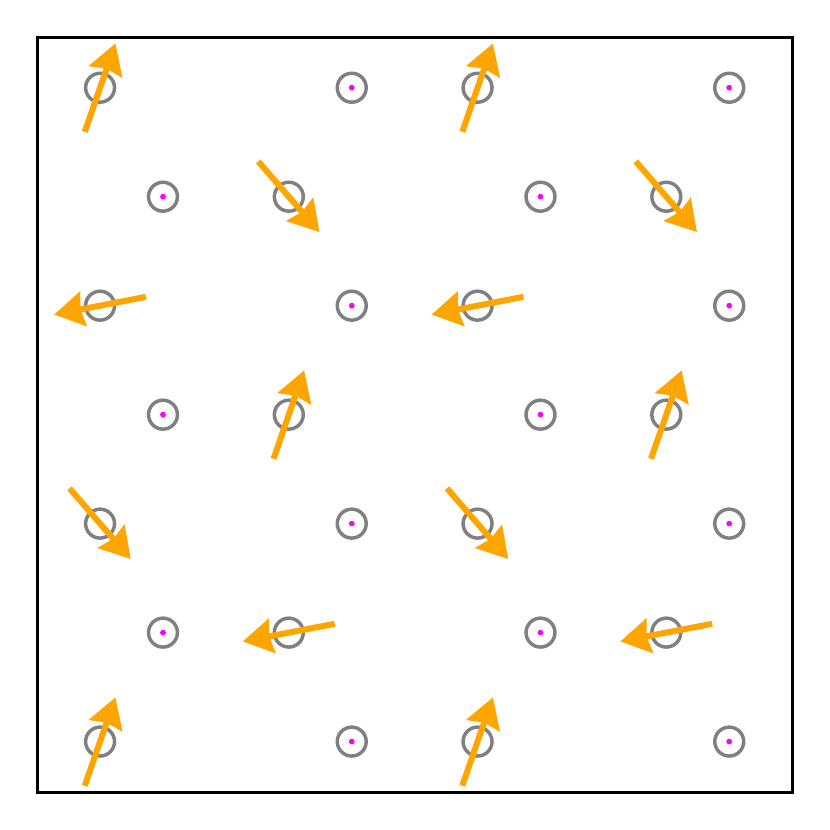}\\
		\includegraphics[width=1\columnwidth]{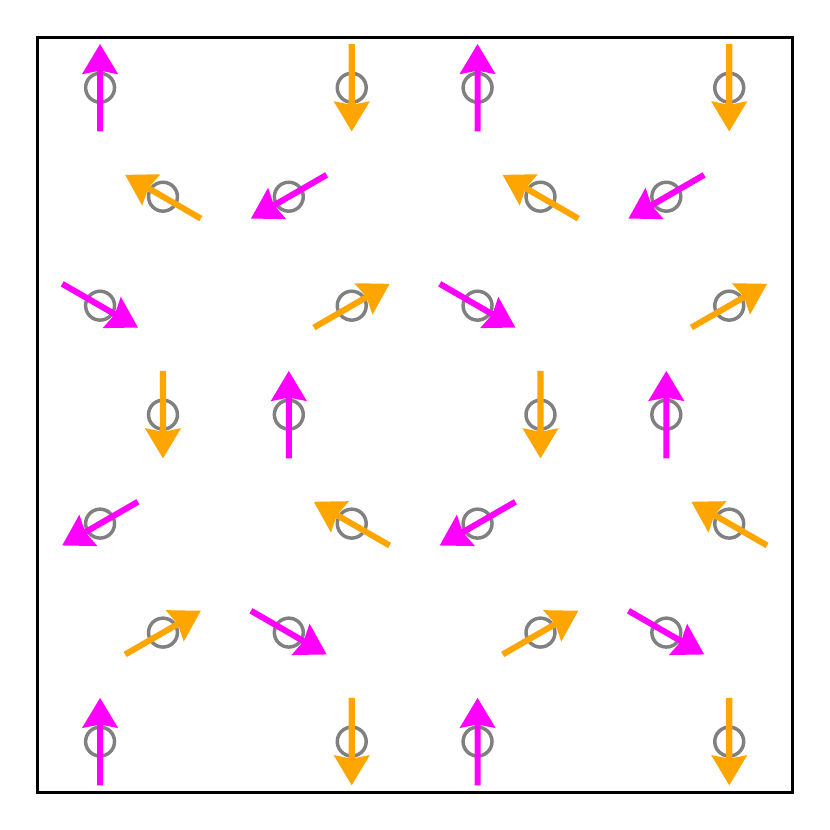}
	\end{minipage}
	%}
	&
	\begin{minipage}[c]{.30\columnwidth}
	\includegraphics[width=1\columnwidth]{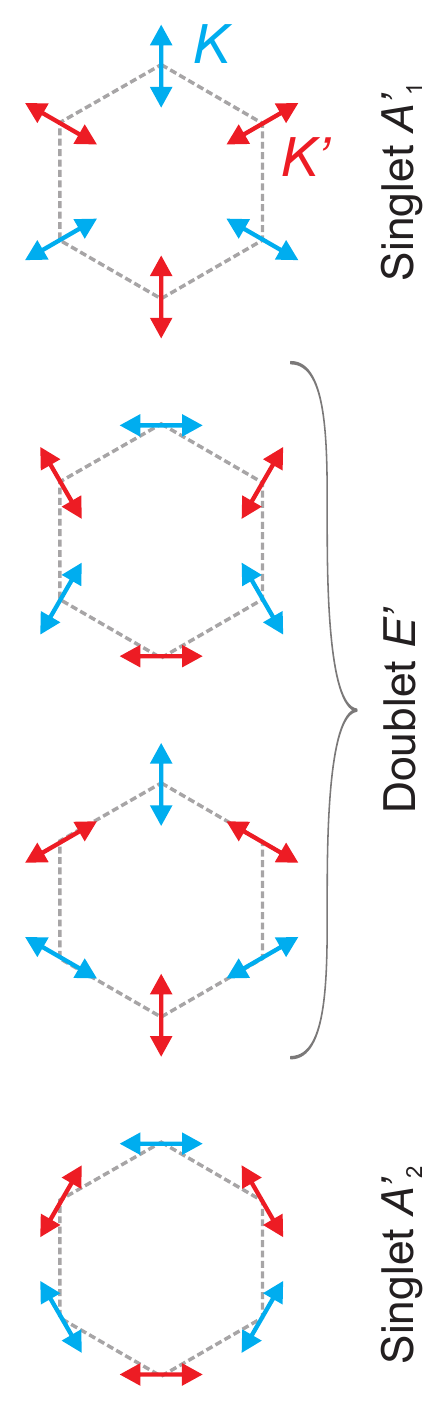} \end{minipage}
		&
	 \end{tabular}
	 }\end{center} \caption{Eigenmodes of the
	honeycomb plasmonic lattice at the $\Kp$-point.  (a) Real-space electric
	dipole polarizations of the nanoparticles (circles) corresponding to
	two singlet modes and a doublet mode, at a specific time.  The dipole
	polarizations depicted by orange and magenta arrows evolve in time
	rotating clockwise and counterclockwise, respectively, for the $\Kp$
	mode, and in the opposite directions for the $\Kp$' mode. (b) Fourier
	transform of the dipole polarizations in the corresponding eigenmodes.
	{(c) Band structure of the empty lattice
	model, that is, as given by diffracted orders of a periodic structure
	without the effect of the localized plasmonic resonance of the
	nanoparticles, with the studied $\Kp$-point highlighted. (d) The first
	Brillouin zone (green area) of the honeycomb reciprocal lattice and its
	high symmetry points, $a$ is the lattice
	constant.
    (e) Singular values \changed{(SV)} of the symmetry-adapted scattering problem
	at the $\Kp$-point, whose minima give the mode energies, as function
	of energy. The color shows the results of projection of the corresponding
	eigenmodes on the singlets and doublets obtained by group theory
	(\removed{orange:} \changed{the} singlet \TheSinglet~\changed{that was
	found to lase experimentally is shown in orange, the other singlet
	$A_2'$ in green}, \removed{blue:} \changed{and the} doublet $E'$
	\changed{in blue}), the energies are marked by ticks.}}
\label{fig:eigenmodes theory} \end{figure}

A single unit cell of the reciprocal lattice of our system contains six high
symmetry points (Fig.~\ref{fig:eigenmodes theory}(\removed{c}\changed{d})): one
$\Gammap$-point with $D_6$ point symmetry, as well as two $\Kp$-points with
$D_3$ and three $\Mp$-points with $D_2$ point symmetries. The $\Kp$-points are
mutually related by parity inversion symmetry.  Whenever the distinction
between the two $\Kp$-points is relevant, we label the other one as $\Kp'$.  To
a large extent, group theory determines the properties of the eigenmodes
supported at the high-symmetry points.  As the reciprocal lattice has $D_3$
point group symmetry around the $\Kp$-points, the $\Kp$-point modes must
constitute irreducible representations of the $D_3$ group.  Using standard
group-theoretical reduction methods \cite{dixon_computing_1970}, we can
determine for instance the electric dipole polarizations of the nanoparticles
in the respective modes.  The irreducible representations of $D_3$ are either
one- or two-dimensional, so the eigenmodes are, apart from accidental
degeneracies, either non-degenerate (“singlets", 1D representation) or doubly
degenerate (“doublets", 2D representation). Six dispersion branches meet at the
$\Kp$-point (see \arxivonly{section \ref{sm:do} of }Supplemental Material), and the
eigenmodes constitute two singlets and two doublets.

Fig.~\ref{fig:eigenmodes theory}(a) shows the admissible patterns of nontrivial
nanoparticle dipole polarizations in the \removed{$\vect{E}$-in-plane }\changed{in-plane-$\vect{E}$}
case for the singlets and one doublet.  Any linear combination of the depicted doublet
states is possible as well.  Fig.~\ref{fig:eigenmodes theory}(b) shows spatial
Fourier transforms of these patterns, corresponding to the polarizations of the
far-field beams escaping the array.

In real space, {the magenta color in
Fig.~\ref{fig:eigenmodes theory} means clockwise rotating electric dipole
polarizations while \removed{at }orange means the dipoles rotate
counterclockwise for all $\Kp$-modes.}
For $\Kp'$-modes, the polarization
rotation directions are reversed.  If the system is excited simultaneously in
the $\Kp$ and corresponding $\Kp'$ state\changed{s} with the same intensities, the
polarizations will, instead of rotating, oscillate in a linear direction, with
the exact direction depending on the relative phase between the $\Kp$ and
$\Kp'$ modes.  This will be important in analyzing the experimental real-space
images.

To characterize the lasing action, we perform angle, energy, polarization and
position resolved emission measurements.  Above a critical pump threshold, the
sample exhibits an intense and narrow emission peak at 1.426~eV and
$k_y\sim4.25\times10^6\textnormal{ m}^{-1}$ (corresponding to an angle of
{$35^{\circ}\pm0.4^{\circ}$} with respect to the sample normal), see
Fig.~\ref{Fig3}\changed{(a–c)}.  The emission intensity and mode line width as a function of
pump fluence is shown in Fig.~\ref{Fig3}(b). 
Over three orders of magnitude increase in emission intensity can be seen
upon the onset of lasing{, typical for nanoparticle arrays with small
spontaneous emission \changed{coupling} to the lasing mode 
(small $\beta$-factor~\cite{Bjork1991})~\cite{Wu2016,Hakala2017,ramezani_plasmon-exciton-polariton_2017,Rekola2018,daskalakis_ultrafast_2018}.}
Increased temporal coherence due to lasing is evident from
the line width of the emission (2 meV), which is well below the natural
line width of the SLR mode at the $\Kp$-point ($\sim20$~meV).  {The 2 meV
line width is smaller than those in
\cite{Zhou2013,schokker_lasing_2014,Yang2015,Wu2016,Zhang2015,ramezani_plasmon-exciton-polariton_2017}
(3.6–27 meV), but larger than the values 0.26–1.5 meV in
\cite{Rekola2018,Wang2017a,Hakala2017,daskalakis_ultrafast_2018,Meng2014}.} 

\begin{figure}[h] \centering
	\includegraphics[width=0.95\columnwidth]{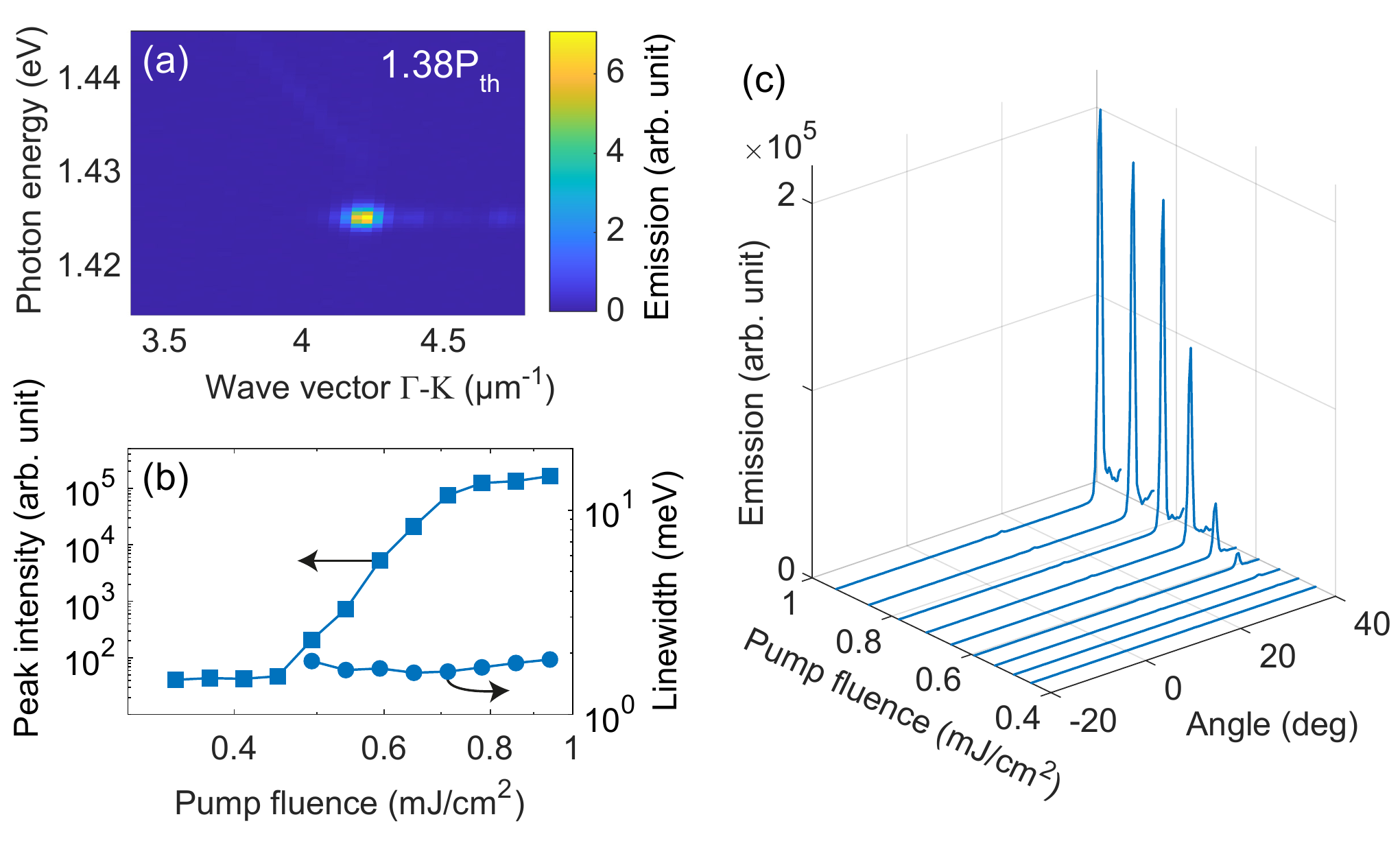}
	\caption{(a) Measured emission spectra of a honeycomb lattice with
	$P=1.38P_\textnormal{th}$, where $P_\textnormal{th}=0.47$~mJ/cm$^2$ is
	the threshold pump fluence for the K-point lasing mode (particle
	distance $p=576\,\mathrm{nm}$ and diameter $d=100\,\mathrm{nm}$).  (b)
	The mode output power (squares) and the line width (circles) at the
	K-point angle ($35^{\circ}$ {$\pm0.4^{\circ}$}) as a function of pump
	fluence. Note that due to low intensity, we cannot determine the line
	width at pump {fluences} below the threshold, for below threshold
	emission, see\arxivonly{ Fig. \ref{smfig:emission_below_threshold} in} Supplemental Material.  (c) The emission intensity as a
	function of angle at the $\Kp$-point energy ($\sim$ 1.426~eV) with
	several pump {fluences}.} \label{Fig3}
\end{figure}

In Fig.~\ref{Fig4}(a), we show the angle resolved emission of the sample. The system
exhibits lasing at six specific angles that correspond to three $\Kp$ and three $\Kp'$
points of the lattice. We measure the polarization properties of each point by
recording the emission intensities with several different linear polarizer
angles. For each point, we recover a typical dipolar emission pattern, however,
the direction of linear polarization is different, see
Fig.  \ref{Fig4}(b). The results match excellently the calculated angular
distributions of linearly polarized light having a polarization along the six
$\Gammap$–$\Kp$ directions (the red dashed lines). We find that the
\TheSinglet~singlet mode has corresponding polarization properties, see
Fig.~\ref{fig:eigenmodes theory}(b).
%\RG{The experimentally measured linear polarization degrees
%$P=(I_\textnormal{max}-I_\textnormal{min})/(I_\textnormal{max}+I_\textnormal{min})$
%are 0.7750, 0.8510, 0.7400, 0.8314, 0.7822 and 0.8138 for the six $\Kp$-points
%respectively (clockwise from the top one). 
{The linear polarization degree
$\rho_\textnormal{L}=(I_\textnormal{max}-I_\textnormal{min})/(I_\textnormal{max}+I_\textnormal{min})$
has an average 
%0.799 
0.8 for the six $\Kp$-points.}

\begin{figure}[h] \centering
	\includegraphics[width=0.95\columnwidth]{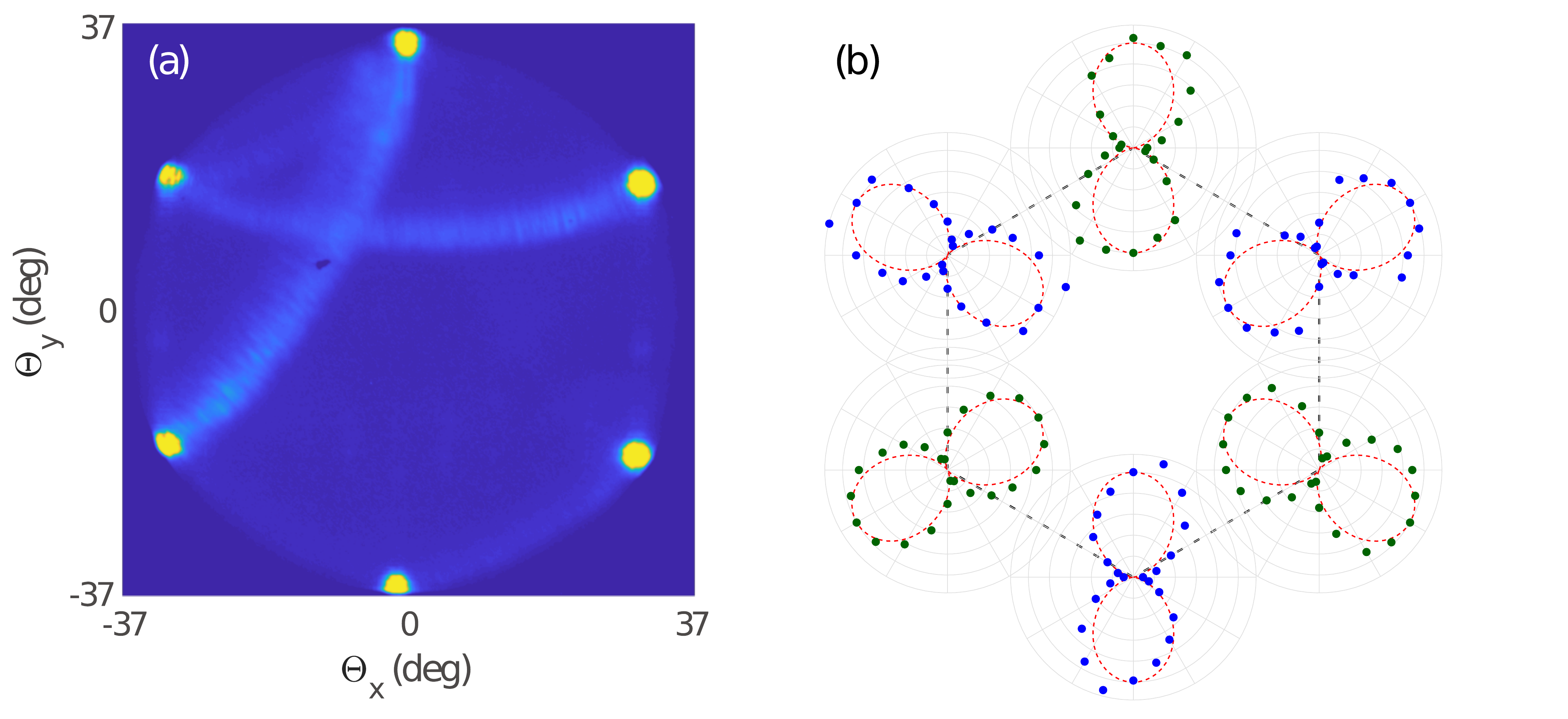}
	\caption{Lasing mode polarization.  (a) Angle resolved emission of the
	sample without any polarizer in detection. All six $\Kp$-points are
	clearly visible. 	(b) Polar emission intensities at each $\Kp$-point
	in the presence of a linear polarizer. The angles refer to the
	polarizer angles and the radii refer to the measured intensities. The
	red dashed lines are the calculated intensity distributions for
	linearly polarized light (along the $\Gammap$–$\Kp$ directions) passing
through the polarizer at the corresponding angle.} \label{Fig4} \end{figure}

The identification of the lasing mode as the singlet \TheSinglet~can be further
confirmed by analyzing the real-space images with variously oriented
polarization filters at the output.  While the dipole polarization directions
of the nanoparticles cannot be measured directly, we can estimate them using
the spatial intensity variations due to wave interference in case of different
filter orientations. T\removed{he clarity of t}he intensity variations should be most
clear in the case where the system lases in the $\Kp$ and $\Kp'$ modes
simultaneously, with a fixed (modulo $\pi/3$) relative phase such that the
dipoles are oriented as in Fig.~\ref{fig:eigenmodes theory}(a).  If the system
lases only in one of the $\Kp$ or $\Kp'$ modes, or if the relative phase is
random, the real-space intensity distribution should become more uniform due to
time averaging (see \arxivonly{section \ref{smfig:phase} of }Supplemental
Material).

Fig.~\ref{fig:real-space samples} shows an image of a small piece of the array
for three choices of polarization filters for the lasing emission, 
with the predicted intensities and nanoparticle electric dipole polarizations
of the singlet mode \TheSinglet~for the ideal, namely zero phase-difference
combination of the $\Kp$ and $\Kp'$ modes, as defined in
Fig.~\ref{fig:eigenmodes theory}(a) (cases with other polarization filter
orientations {and details of the theoretical predictions} are shown in
\arxivonly{section \ref{sm:rs} of }Supplemental Material). The intensity maxima
appear at the places where the surrounding adjacent dipoles, or their
projections according to the
polarization filter orientation, have the same or similar directions and
therefore interfere constructively. 
Comparing the real-space images with dipole orientations predicted for the
other modes ($A_2'$ and the doublet $E'$) results in inconsistencies (for
details, see Supplemental Material\arxivonly{, section \ref{sm:rs}}).
This confirms that the system indeed
lases in the singlet mode \TheSinglet. The intensity variations in the observed
patterns show that the system lases in the $\Kp$ and $\Kp'$ singlet
\TheSinglet~modes simultaneously, with comparable intensities and with a fixed,
or at least strongly correlated, relative phase. The existence of interference
patterns {over the whole sample}, furthermore, proves the spatial coherence
of the observed lasing.  
Since the $\Kp$-point of our system corresponds to the crossing of
diffractive orders in three directions with $120^{\circ}$ angles between them,
the feedback in the lasing action is two dimensional{, different from one
dimensional DFB lasing~\cite{Kogelnik1972} in nanoparticle arrays
\cite{Wang2017,Wang2017a,Rekola2018,Hakala2017}}. This is
reflected in the non-trivial 2D polarization patterns. 

{DFB-type lasing typically occurs at a band edge or an extremum of the
dispersion because zero group velocity enables feedback. Both the measured and
simulated dispersions (Fig.~\ref{Fig1}(a)) show crossings of the modes at the
$\Kp$-point, without any visible gap and zero group velocity point. Why does a
mode of a certain symmetry (the \TheSinglet~singlet) lase, if the $\Kp$-point
apparently has a degeneracy of several modes? 
{To answer this we computed the energies of the eigenmodes
using symmetry-adapted $T$-matrix simulations} (for details, see%
\arxivonly{ sections \ref{sm:tmatrix}–\ref{sm:symmetries} of} Supplemental Material).
\changed{Fig.~\ref{fig:eigenmodes theory}(e) shows that indeed there is a 
difference in the energies of the  \TheSinglet~singlet and the $E'$
doublet near the $\Kp$-point. This band gap means that the singlet
\TheSinglet~has an energy minimum at the $\Kp$-point, which explains why lasing
is possible in this mode. The narrower peak for \TheSinglet~compared to that
for $E'$ indicates higher quality factor, making the former mode more amenable
for lasing. The $A_2'$ singlet mode seems almost degenerate with \TheSinglet{} but the
resonance is a bit weaker (slightly smaller dip in Fig.~\ref{fig:eigenmodes
theory}(e); see \arxivonly{Fig.~\ref{smfig:dispersion detail} of }Supplemental Material 
for a larger picture)
The energy difference between \TheSinglet~and $E'$ is only 3.2 meV,
smaller than the natural linewidth of the SLR mode around 20 meV, which
explains why the gap is not visible in the dispersions. On the other hand, the
lasing emission has 2 meV linewidth, similar to the scale of the band gap.}
\removed{Fig.~\ref{fig:eigenmodes theory}(e) shows that indeed there is a
difference in the energies of the  \TheSinglet~singlet and the $E'$ doublet
near the $\Kp$-point, albeit small, about 3.2 meV which is smaller than the
natural linewidth of the SLR mode around 20 meV which explains why the gap is
not visible in the dispersions. On the other hand, the lasing emission has 2
meV linewidth, similar to the scale of the band gap.  The small band gap means
that the singlet
\TheSinglet~has an energy minimum at the $\Kp$-point, which explains why lasing
is possible in this mode. The narrower peak for \TheSinglet~compared to that
for $E'$ indicates higher quality factor, making the former mode more amenable
for lasing. The $A_2'$ singlet mode seems degenerate with \TheSinglet~but the
resonance is a bit weaker (slightly smaller dip in
Fig.~\ref{fig:eigenmodes theory}(e).}}

\begin{figure} 
	\begin{center}
	\altcombinedimage{real-space-samples.pdf}{
		\includegraphics[width=0.32\columnwidth]{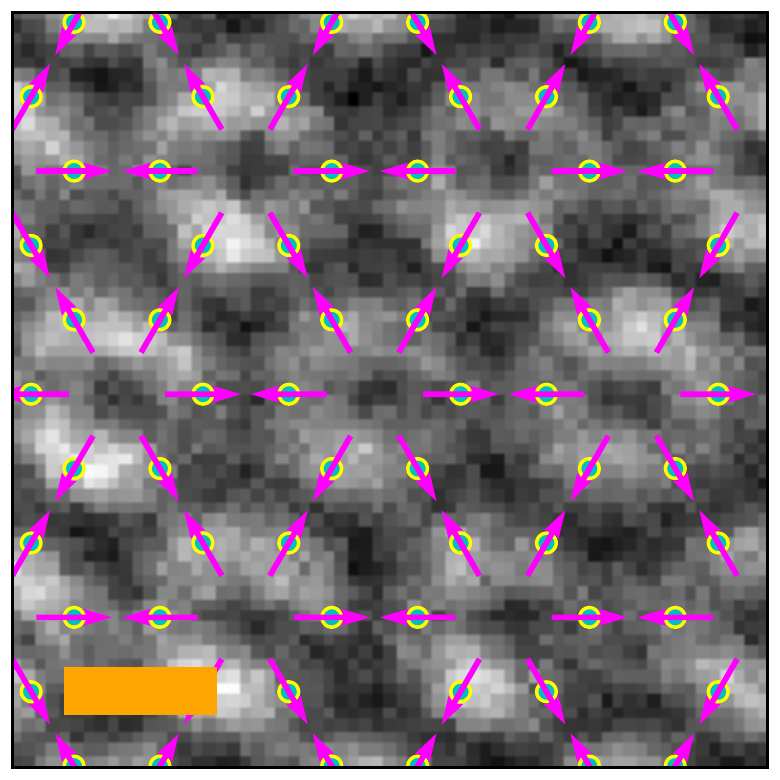}%
		\includegraphics[width=0.32\columnwidth]{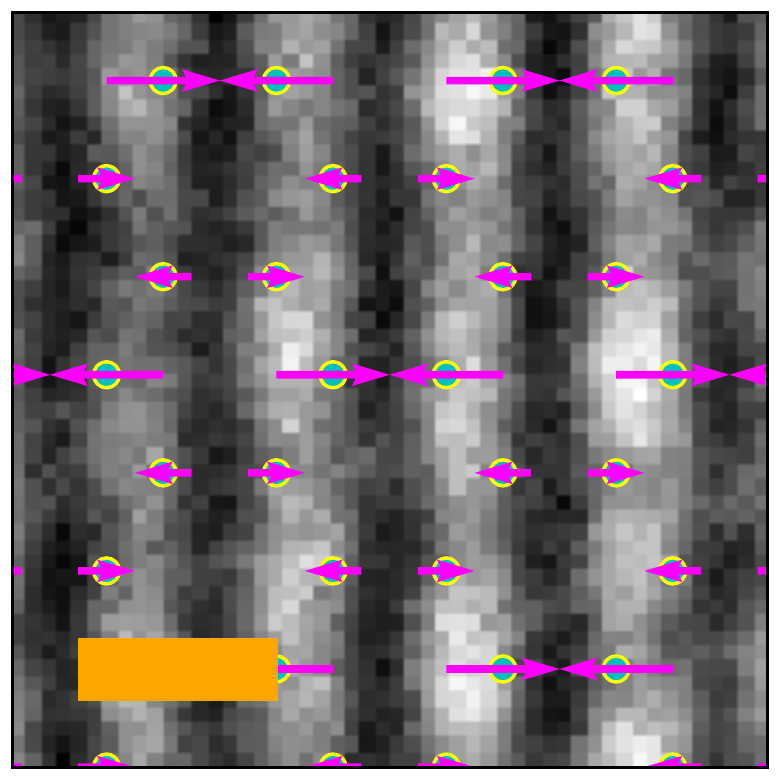}%
		\includegraphics[width=0.32\columnwidth]{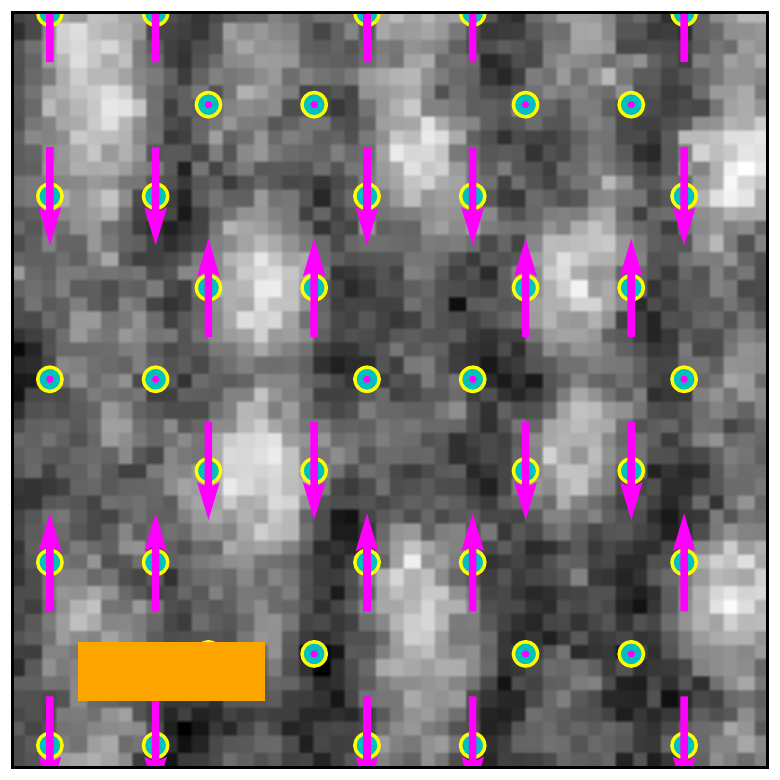}
		\\
		\includegraphics[width=0.32\columnwidth]{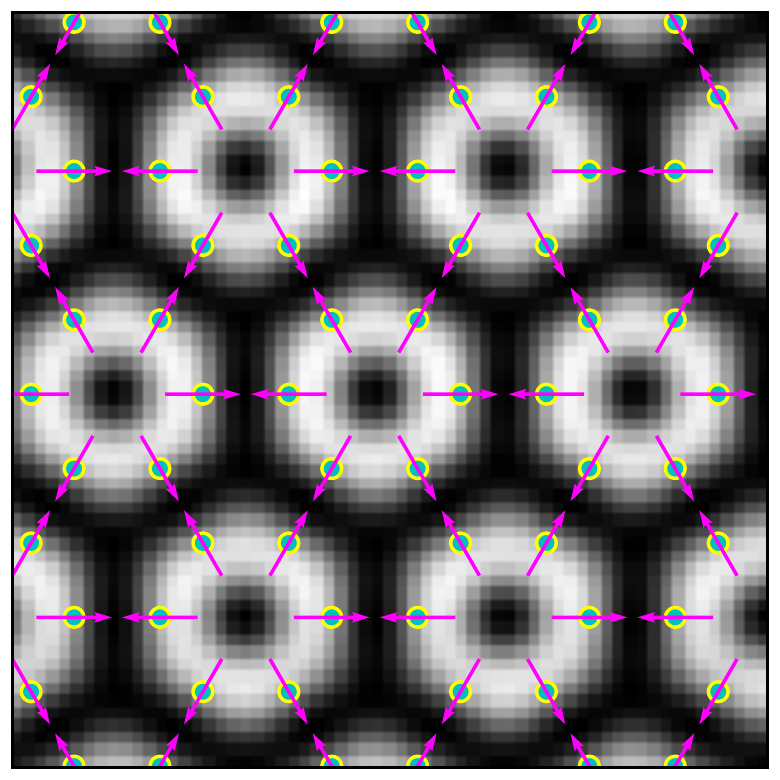}%
		\includegraphics[width=0.32\columnwidth]{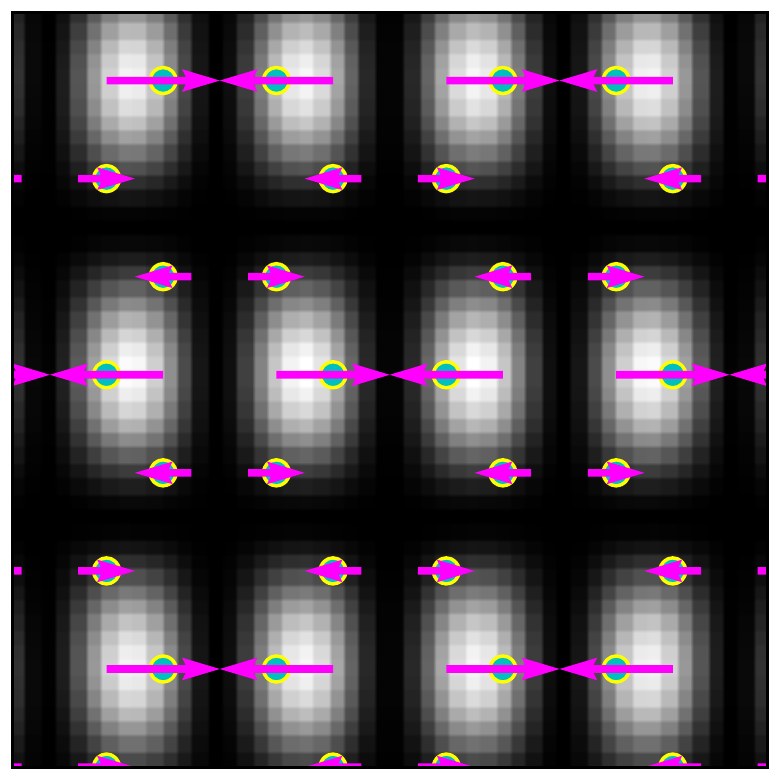}%
		\includegraphics[width=0.32\columnwidth]{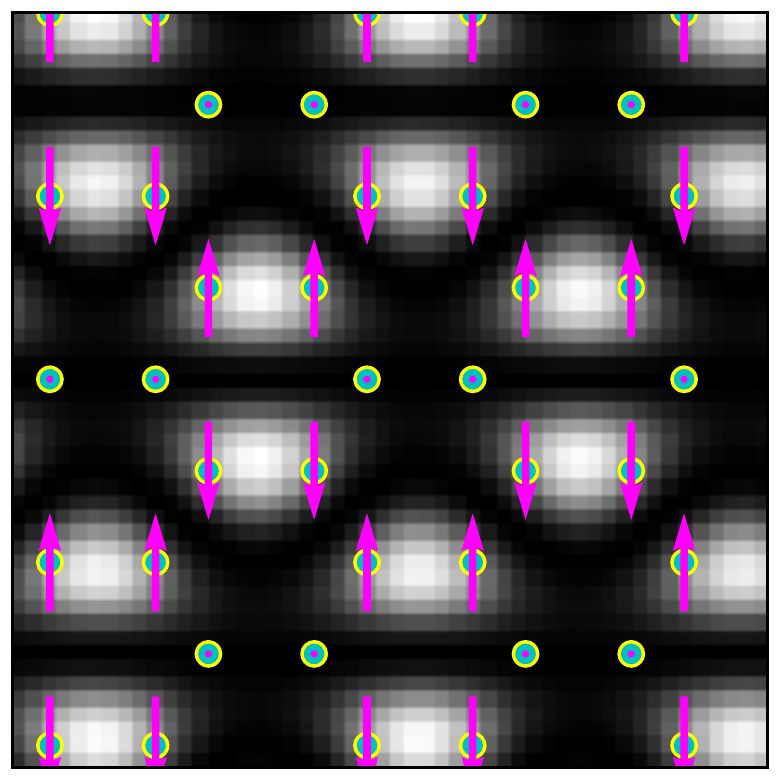}
		\\
		\begin{tabular}{p{0.3\columnwidth}p{0.3\columnwidth}p{0.3\columnwidth}}
		{\centering No filter \par} & {\centering Filter
		$\leftrightarrow$ \par} & {\centering Filter $\updownarrow$
		\par} 
		\end{tabular} 
	}
	\end{center}
	\vspace*{-0.8cm}
	\caption{Upper row: examples of real-space images for different
		polarization filters (no filter, horizontal, vertical) used for
		the analysis of the lasing mode.  In each case, the expected
		positions of the nanoparticles {(small yellow-cyan circles)} and
		the dipole polarizations (arrows) of the singlet mode
		\TheSinglet~for the ideal (zero) $\Kp$—$\Kp'$ relative phase
		are depicted, projected to the corresponding filter direction.
		Lower row: theoretical prediction for the intensities for the
		ideal $\Kp$—$\Kp'$ relative phase.  The scale bar length is
		$1\,\mathrm{\mu m}$.  For images over larger areas, see \arxivonly{section
		\ref{sm:rs} of} Supplemental Material.	 \label{fig:real-space samples}}
\end{figure}

In summary, we have observed lasing action at the $\Kp$ and $\Kp$' points
of a honeycomb plasmonic lattice. Both the polarization of the six output beams
and the real space interference patterns provide distinct features that, when
combined with the group theory description, reveal the lasing mode as the
singlet \TheSinglet. Analysis of the $T$-matrix simulation results using the
group theory eigenmodes showed that the singlet \TheSinglet~has an energy
minimum at the K-point, which enables the feedback necessary for lasing.
\diffnote{DIVISION INTO A NEW PARAGRAPH REMOVED HERE.} {Our results demonstrate the
potential of plasmonic nanoparticle array systems for tailoring the
polarization and beam direction of laser output by the lattice geometry. The
tunability of the beam direction (here $\sim 35^{\circ}$) can be used for
bringing the beam close to the in-plane direction \removed{to enable on-chip
planar integration}. \changed{If realized in a less lossy platform, this could
enable on-chip planar integration.} 

\diffnote{A NEW PARAGRAPH START INTRODUCED HERE.}
Our study gives a promising starting point for
investigations of topological photonics and
lasing~\cite{bahari_nonreciprocal_2017,Khanikaev2017,ozawa_topological_2018,harari_topological_2018,bandres_topological_2018,Lu2014,Lu2016,Haldane2008,Raghu2008}
in radiatively coupled systems. Plasmonic nanoparticle array lasers offer a
unique combination of easy fabrication, room temperature operation, ultrafast
speeds, long-range radiative coupling, and strong coupling to emitters (the
gain
medium)~\cite{Vakevainen2014,Torma2015,ramezani_plasmon-exciton-polariton_2017}.
Radiatively coupled systems offer topological phenomena different from
tight-binding models~\cite{Pocock2018}. 
\changed{Arrays
of magnetic nanoparticles have been realized~\cite{Kataja2015}, and the
magnetization of nanoparticles could be used for opening topological gaps at
the high-symmetry points where we have shown lasing. Time reversal symmetry
breaking is one of the main mechanisms leading to topologically non-trivial
systems but topological gaps based on magnetic
materials~\cite{Haldane2008,Raghu2008} are extremely small at optical
frequencies~\cite{bahari_nonreciprocal_2017}. The polarization and
interference analysis demonstrated here will be invaluable in identifying
topological modes even when related gaps would be small. Remarkably, the lasing
action is stable despite a small gap, which
is promising concerning topological lasing relying on small topological gaps.}

\removed{Time reversal symmetry breaking is one of the main mechanisms leading to
topologically non-trivial systems and could be realized by magnetic materials. 
But the resulting topological gaps are
extremely small at optical frequencies. Arrays
of magnetic nanoparticles have been realized, and the
magnetization of nanoparticles could be used for opening topological gaps at
the high-symmetry points where we have shown lasing. The polarization and
interference analysis demonstrated here will be invaluable in identifying
topological modes even when related gaps would be small. Remarkably, the lasing
action is stable despite a small gap, which is promising concerning topological 
lasing relying on small topological gaps.} 

\begin{acknowledgments} 
This work was supported by the Academy of Finland
through its Centres of Excellence Programme (Projects No.\ 284621, No.\ 303351,
No.\ 307419) and by the European Research Council (Grant No.
ERC-2013-AdG-340748-CODE).  This work benefited from discussions and visits
within the COST Action MP1403 Nanoscale Quantum Optics, supported by COST
(European Cooperation in Science and Technology).  Part of the research was
performed at the Micronova Nanofabrication Centre, supported by Aalto
University. Computing resources were provided by the Triton cluster at Aalto
University. The authors thank Matthias Saba, Ortwin Hess, Tero Heikkilä and
Heikki Rekola for fruitful discussions.
\end{acknowledgments}

%\bibliography{Refs,hexarray-theory,Refs_misc}


\clearpage
\onecolumngrid
\vspace{5ex}
\begin{center}
	\textbf{\large Lasing at the $K$-points of a honeycomb plasmonic lattice\\\vspace{5pt}
	 Supplemental Material}
\end{center}
\vspace{5ex}
\twocolumngrid

\setcounter{equation}{0}
\setcounter{figure}{0}
\setcounter{table}{0}
\setcounter{page}{1}
\makeatletter

\renewcommand{\thepage}{S\arabic{page}}
\renewcommand{\theequation}{S\arabic{equation}}
\renewcommand{\thefigure}{S\arabic{figure}}
\renewcommand{\bibnumfmt}[1]{[S#1]}
\renewcommand{\citenumfont}[1]{S#1}

%\clearpage
\author{R.\ Guo$^1$}
\author{M.\ Nečada$^1$}
\author{T.K.\ Hakala$^{1,2}$}
\author{A.I.\ Väkeväinen$^1$}
\author{P.\ Törmä$^1$}
\email{paivi.torma@aalto.fi}
\affiliation{$^1$Department of Applied Physics, Aalto
University, FI-00076 Aalto, Finland \\
{$^2$Institute of Photonics, University of Eastern Finland, P.O.Box 111, FI-80101 Joensuu, Finland}}

\title{Lasing at the $\Kp$-points of a honeycomb plasmonic lattice\\Supplemental material}

\maketitle

\section{Experimental Methods}\label{sm:methods}

\subsection{Sample fabrication} 

Honeycomb lattices of cylindrical gold nanoparticles (diameter 100 nm, height 50 nm) were
fabricated on borosilicate substrate with electron beam lithography (Vistec
EPBG5000pES, {acceleration voltage: 100kV}). Two nanometers of titanium was
deposited prior to gold deposition to provide an adhesive layer. The overall
size of the array was 
$100\times100 \mathrm{\mu m^2}$.

\subsection{Gain medium}

{The gain medium used was dye IR-792 perchlorate purchased from
Sigma-Aldrich. The dye molecule was dissolved into 1:2 (dimethyl
sulfoxide):(benzyl alcohol) solvent with a concentration of 25 mM. Figure
\ref{smfig:ir792emission} shows the emission spectrum of the dye solvent with
the same concentration.  Amplified spontaneous emission and lasing of IR-792
have been
reported previously in systems of dye-doped polymer thin film
\cite{Thompson2004} and of edge-pumped plasmonic lattice
\cite{hakala_bec_2018}.}

\begin{figure}[b]
\begin{center}
\includegraphics[width=.45\textwidth]{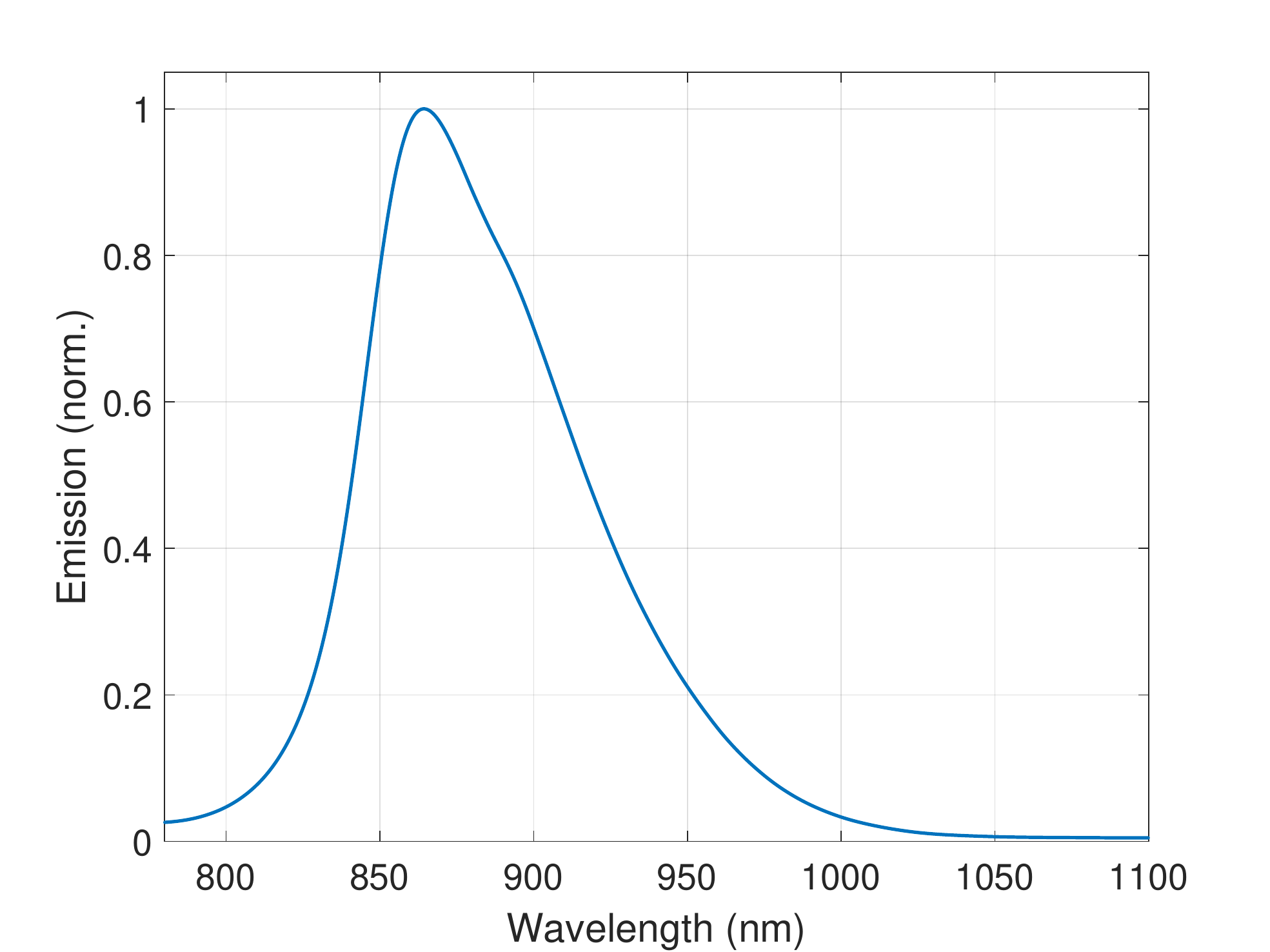}
\end{center}
\caption{{Measured emission spectrum of the IR-792 dye.} 
}
\label{smfig:ir792emission}
\end{figure}

\begin{figure*} 
\begin{center}
\includegraphics[width=.95\textwidth]{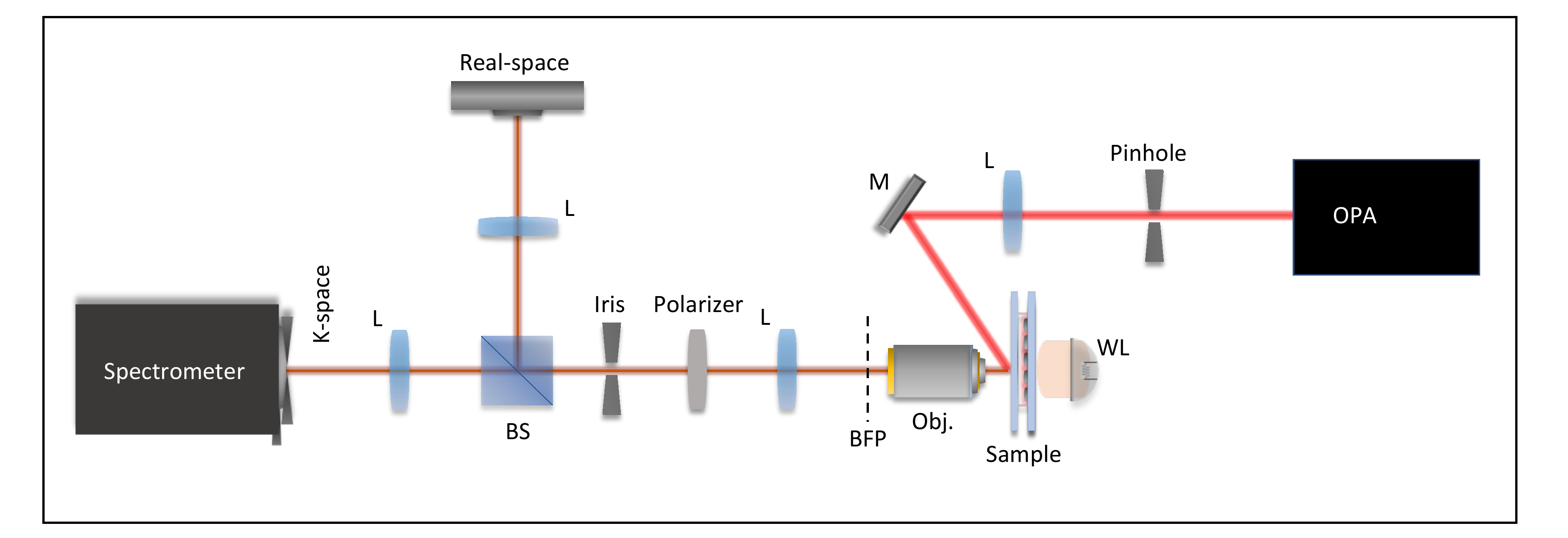}
\end{center}
\caption{{Measurement setup. L stands for lens, M for mirror, BS for beam
	splitter, BF\MMNmark{P} for back-focal plane, Obj. for objective, WL
	for white light source and OPA for optical parametric amplifier. The
	OPA is used to tune the pump wavelength, here we used 750~nm.} 
%{Was it 750~nm in these experiments???}. {Yes, at least that is given in the
	%main text.}
}
\label{smfig:expschematic}
\end{figure*}

\subsection{Measurement setup}

{A schematic of the measurement setup is presented in Figure~\ref{smfig:expschematic}}.
The back focal plane of a $40 \times 0.6$ NA microscope objective was
focused to the entrance slit of a spectrometer {with a focal length of
500 mm and a spectral resolution of $\sim0.5$ nm}. The angle-resolved
extinction spectra (the dispersion) were obtained by focusing light from a
halogen lamp onto the sample. \removed{The sample slide was tilted $10^\circ$
to allow for large angles to be measured.}%
\changed{The sample substrate was placed in a $10^\circ$ tilted stage in order
to collect light asymmetrically from higher angles than the objective NA would
allow at normal incidence.} The measured spectra (namely, the images from the
spectrometer CCD) were further
calibrated by the diffraction pattern from a 300 lines/mm transmission
grating. {The metal nanoparticle array was fabricated on top of a glass
substrate, and for the lasing measurement, a 1~mm thick dye (IR-792)
solution layer (of volume $\sim 300\,\mathrm{\mu \ell}$) was added on top
of the array. The solution layer was sealed between the glass substrate
and a glass superstrate. The 1~mm dye solution layer on top of the
array is much higher than the
extension of the fields related to the nanoparticles, and also high enough not
to create a waveguide mode at the wavenumbers considered. Therefore the exact
alignment of the two glass slides with respect to each other is not essential.}
The dye solution was pumped with a
femto-second laser with $\sim 60^\circ$ incident angle, 750 nm central
wavelength, 100 fs pulse width and 1 kHz repetition frequency {at room
temperature. The laser spot size on the sample is $\sim 4.4 \times 10^5
\mathrm{\mu m^2}$.} 
The real and back focal plane images of the lasing action
were taken by focusing them onto two separate 2D CCD cameras. {The polarizer
used in the polarization measurements was Thorlabs nanoparticle linear film
polarizer (model LPVIS100-MP2) which has an extinction ratio of $>10^6:1$ in the
wavelength range of interest.}

\subsection{Lasing results}

{Figure~\ref{smfig:thresholdcurvecomparison} shows a comparison of lasing
threshold curves for the peak intensities and integrated intensities under the
lasing peak. Figure~\ref{smfig:thresholddoublelog} shows the data of Fig.~\ref{Fig3}(c)
of the main text in double logarithmic scale.}
{Figure~\ref{smfig:emission_below_threshold} shows the measured emission
spectrum below the threshold.} 
%{Is Figure~\ref{smfig:thresholddoublelog} actually in semi-log scale? The same concerns its caption.}

\begin{figure}[h]
\begin{center}
\includegraphics[width=.45\textwidth]{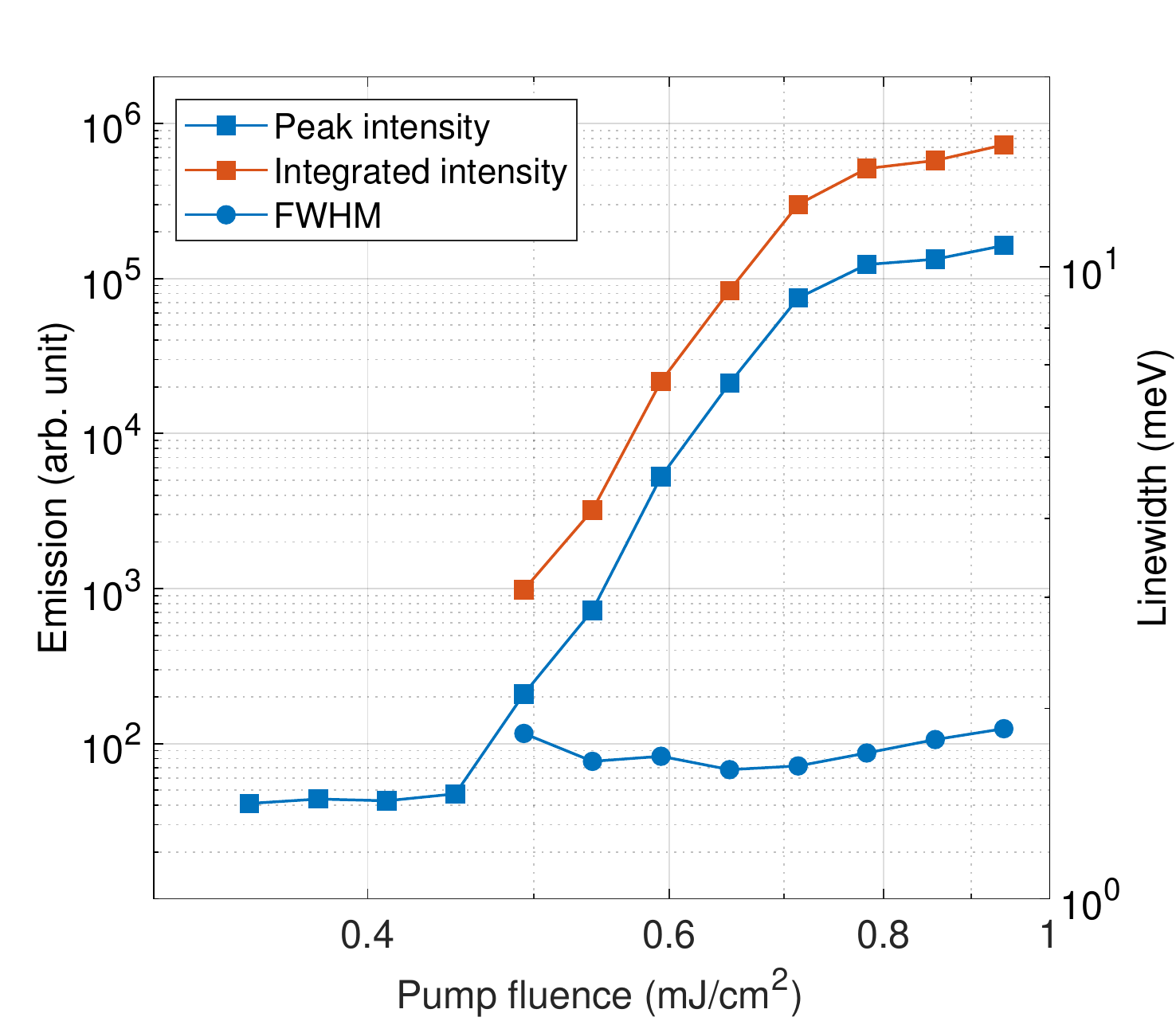}
\end{center}
\caption{{Comparison of the threshold curves with just the peak value (blue)
	and with integrated intensity under the lasing peak. Both show the same
threshold behavior.} 
}
\label{smfig:thresholdcurvecomparison}
\end{figure}

\begin{figure}[h] 
\begin{center}
\includegraphics[width=.45\textwidth]{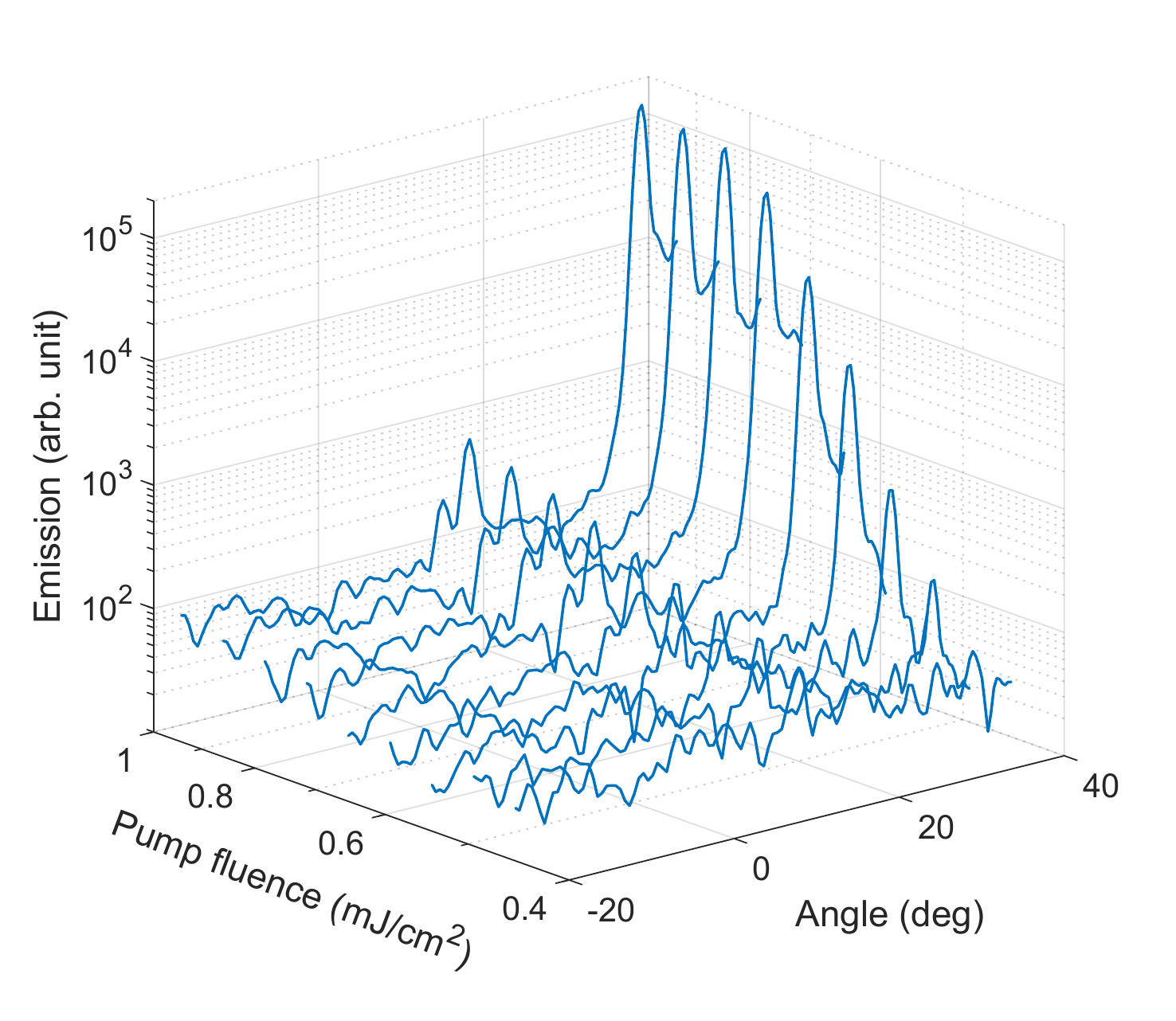}
\end{center}
	\caption{{Data of the Figure \ref{Fig3}(c) of the main text in double logarithmic scale.} 
}
\label{smfig:thresholddoublelog}
\end{figure}

\begin{figure}[h] 
\begin{center}
\includegraphics[width=.35\textwidth]{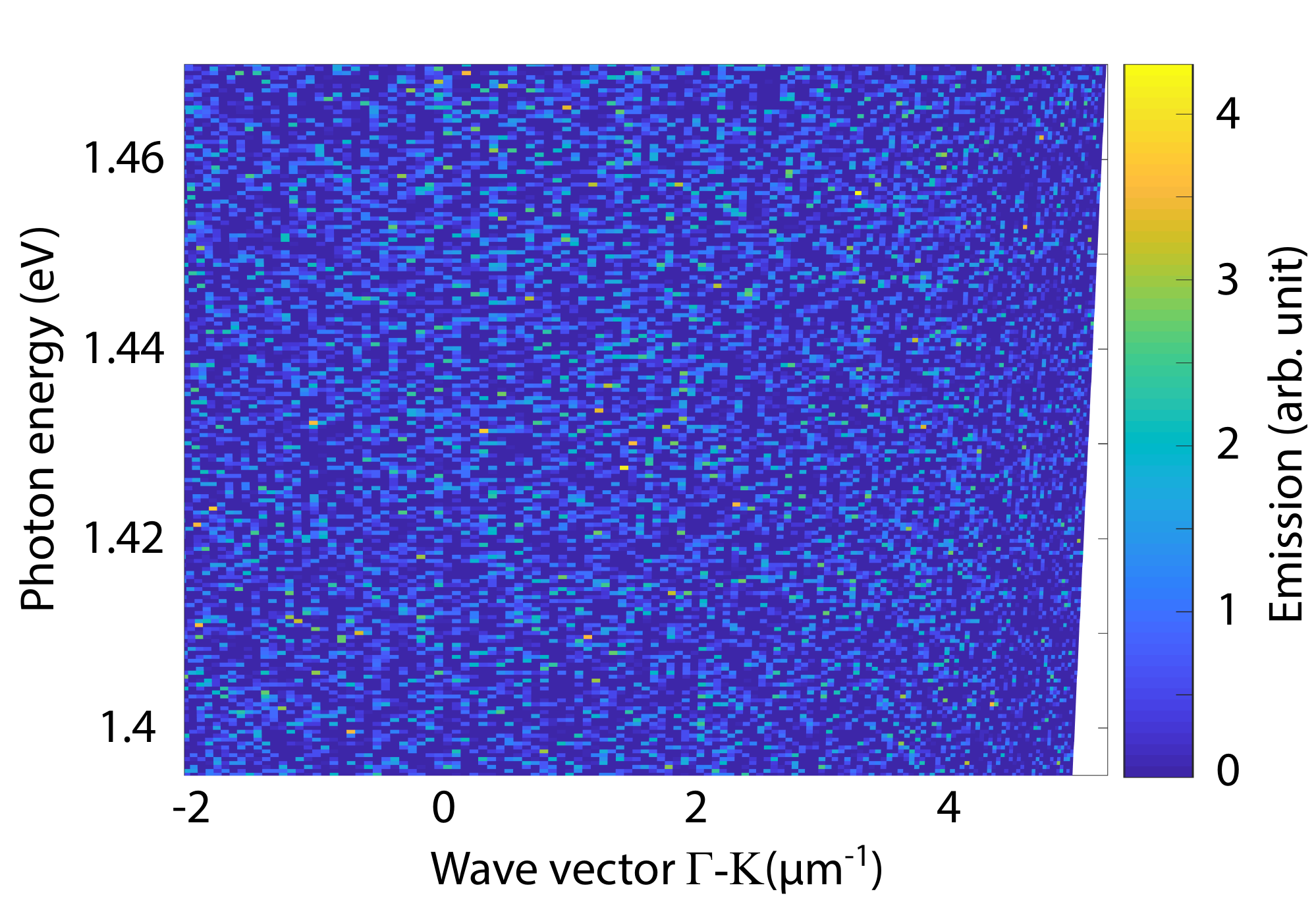}
\end{center}
\caption{{Measured emission spectrum below the threshold ($P=0.75P_\textrm{th}$).} 
}
\label{smfig:emission_below_threshold}
\end{figure}

\section{Diffraction orders and number of modes}\label{sm:do}
The diffraction orders, that is, the empty lattice calculation for our honeycomb
array are shown in Fig.~\ref{smfig:do}. The right panel is the same as shown in
Fig.~\ref{fig:eigenmodes theory}(c) of the main text. The left panel shows a crosscut at the $\Kp$-point
energy, from where one can see that six dispersion branches (in-plane polarized
light cones) meet at the $\Kp$-point. Correspondingly, there will also be six
eigenmodes: two singlets and two doublets as following from decomposition of the
vector space spanned by linearly combining plane waves (or dipole
{polarization} degrees of freedom) into irreducible representations (subject
to the $D_3$ symmetry of given 
$\Kp$-point)~\cite{dresselhaus_group_2008,dixon_computing_1970}.

For background refractive index $1.52$ and $576\,\mathrm{nm}$ spacing between
neighboring nanoparticle centers (i.e. $998\,\mathrm{nm}$ lattice period), 
the third crossing of diffraction orders at the $\Kp$-point happens
at energy $1.44\,\mathrm{eV}$. {The slight difference from the energy of
1.426 eV from Fig.~\ref{Fig3} is most likely due to the real background index
of refraction not having exactly the value of 1.52 used in the simulations.} 

\begin{figure}\begin{center}
	%\fbox{
		\includegraphics[trim=2cm 1cm 0cm 1cm,width=1\columnwidth]{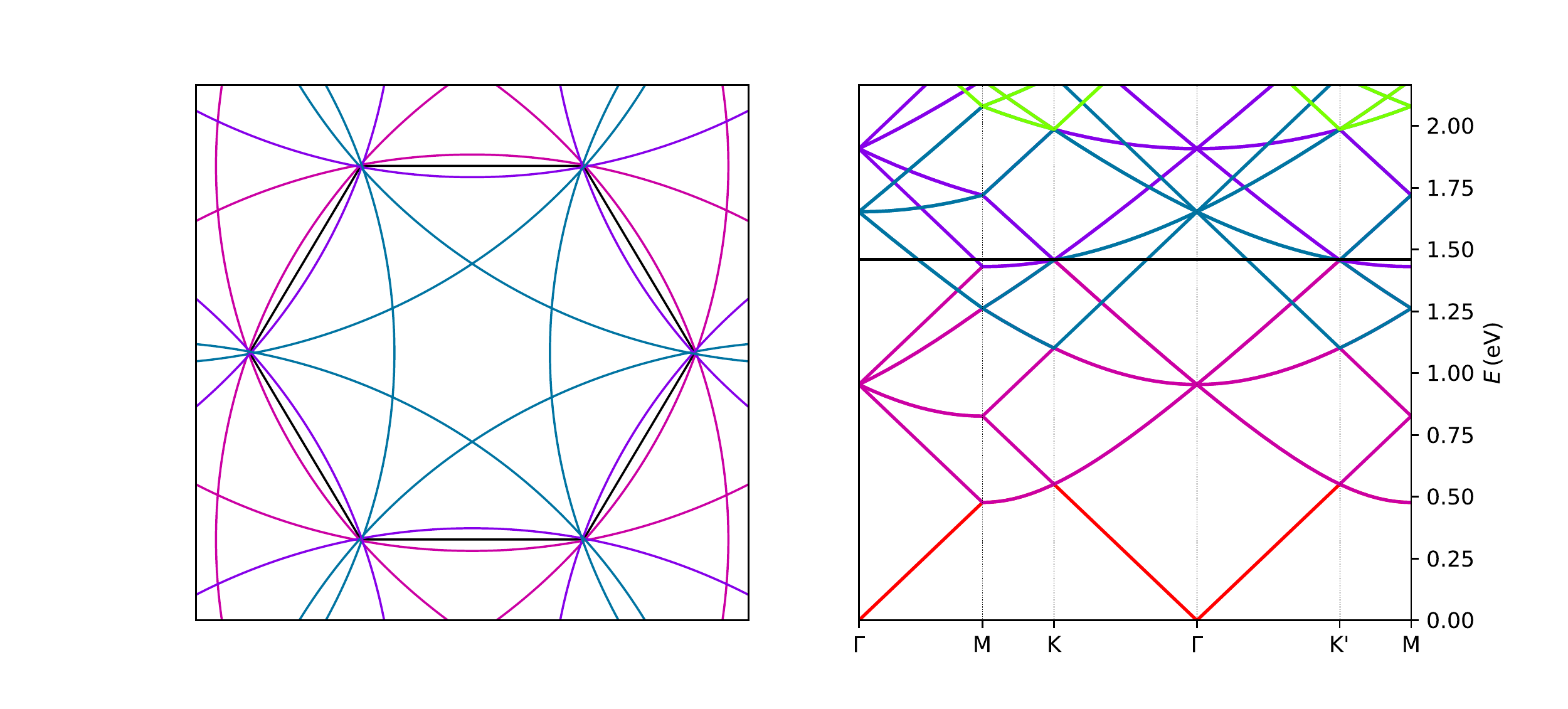}
	%}
		\\
\end{center}
\caption{Diffraction orders of the honeycomb lattice in the empty lattice
description, right along the high symmetry point lines, and left a crosscut at
the energy of the $\Kp$-point of interest in this manuscript. The colors denote
different diffracted orders: red for the
0$^\textrm{th}$, fuchsia for the 1$^\textrm{st}$, navy blue for the
2$^\textrm{nd}$, purple for the 3$^\textrm{rd}$ and green for the
4$^\textrm{th}$.}
\label{smfig:do}
\end{figure}

\section{Determining the lasing modes {in real space}}\label{sm:rs}

{In our experiment, the diameter of a single nanoparticle is much smaller than the
wavelength, hence the nanoparticle can be considered as a monochromatic point source
when lasing. 
Imaged with a sufficient resolution, such source will appear as 
a diffraction pattern rather than
a dot. The exact profile of the pattern will depend on the actual optical setup,
but for practical purposes of our real-space pattern analysis, it can be modeled
as an Airy pattern. In such case, the source $s$ will create electric field 
$\vect{E}_s\left(\vect r'\right)$ at spot $\vect r'$ of the image plane, where 
\begin{equation}
\vect{E}_s\left(\vect r'\right)\propto \vect p_s 
 \frac{ J_1\left(\alpha \abs{\vect r' - \vect{R'}_s}\right)} {\alpha \abs{\vect
	r' - \vect{R'}_s}},
\end{equation}
 $\vect p_s$ being the nanoparticle electric dipole moment,
 $\vect R'_s$ the position of the particle image centre, $\alpha$ an inverse-length
parameter depending on the setup, and $J_1$ the first order regular Bessel function. 
When multiple sources are present, their diffraction patterns will interfere with
each other.
Their respective electric field contributions are
summed up, giving the resulting intensity at the image plane 
$I(\vect r') \propto \abs{\sum_s \vect{E}_s(\vect r')}$.
This is how we obtained the predicted patterns in Fig.~\ref{fig:real-space samples}
of the main text, as well as those mentioned here below.
The exact choice of parameter $\alpha$ does not qualitatively affect the predicted
patterns inside the array as long as the distance between image centres
of two neighboring particles $|\vect R'_{s_1} - \vect R'_{s_2}|$ is well below the radius
of first Airy disk minimum $\approx 1.22/\alpha$ (i.e. if the central circles
of the Airy patterns of neighboring particles overlap), nor are the predicted array
patterns qualitatively changed if the Airy functions are replaced with Gaussian disks.}

{The profiles of the measured diffraction patterns 
will differ from the ideal
Airy or Gaussian patterns (depending on the setup and optical components used)
and are not exactly known, but the respective measured/predicted array patterns will
match at the scale of several unit cells.
At larger scales, however, the total optical path between the source nanoparticles and
and their corresponding image locations on the CCD will differ for different
parts of the array, causing additional phase shifts
in the observed patterns throughout the array.}

\subsection{Large-scale real-space images}
In Fig.~\ref{smfig:wholearrays} we show the same pictures as in Fig.~\ref{fig:real-space samples} of the main
text, but over a larger area (right column) and also covering the whole sample
(left column). The experimentally measured interference patterns extend over the whole sample
size. But the observed pattern sometimes varies throughout the array; {our hypothesis is that 
this is mainly due to the phase shifts depending on the construction of the measurement system
as described above.}
However, any conclusions on this will require further study.
\begin{figure*}
\begin{center}
\includegraphics[width=.8\textwidth]{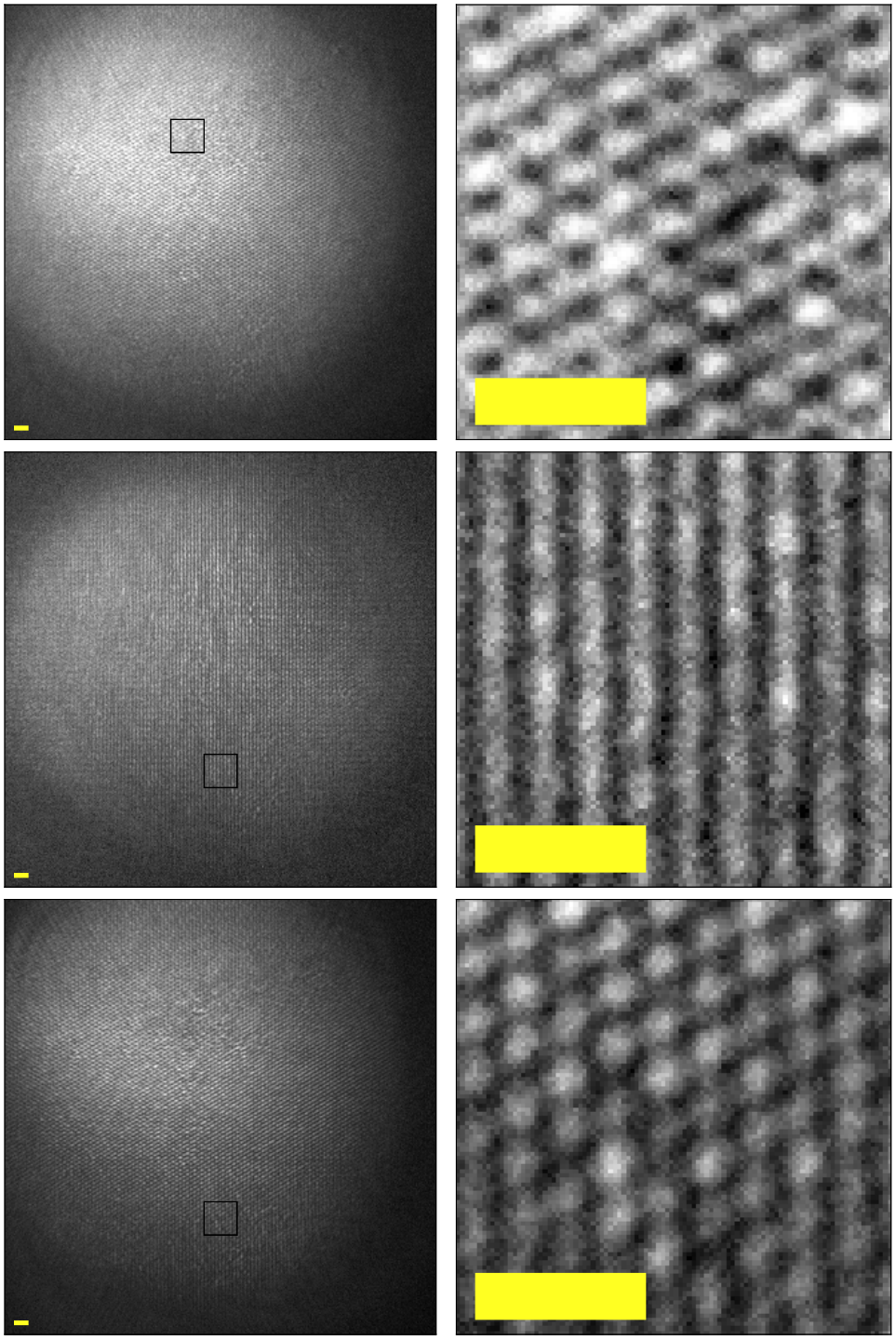}
\end{center}
\caption{Left: real space image of the sample. 
Right: An enlarged image of the position marked by a square in the corresponding
	image on left. The images are for different {polarization} filters,
	top: no filter, middle: horizontal filter, bottom: vertical filter. 
	{The scalebars are $3\,\mathrm{\mu m}$ long.}
}
\label{smfig:wholearrays}
\end{figure*}

\subsection{Comparison of real-space images for different eigenstates}

Fig.~\ref{smfig:comparison} displays a set of measured real space intensity
profiles for different polarization filter orientations, together with the
predicted intensity patterns for the two singlets \TheSinglet{} and $A_2'$, as
well as for the doublet $E'$, with certain superpositions of the doublet states.
The relative phases of the $\Kp$ and $\Kp'$ point lasing modes in the
\TheSinglet{} and $A_2'$ columns are uncorrelated. These pictures demonstrate
that, by inspecting the real space images for multiple values of the emission
detection polarizer filter angle, one can distinguish a certain mode (here the
singlet \TheSinglet) from the other singlet as well as from a combination of the
doublet states. While a single polarizer filter angle result would leave
ambiguity between certain states, a tomographic polarization analysis using
multiple angles leads to unambiguous results: for instance, the \TheSinglet{} and
the first doublet combination produce somewhat similar image for the polarizer
angle $-\pi/2$, but clearly distinct results for the angle $\pi/3$. These images
also show that random, uncorrelated phases between the $\Kp$ and $\Kp'$ lasing
contributions do not produce the precise pattern observed experimentally:
sometimes, a match is found when assuming a constant relative phase between
them, as shown in Fig.~\ref{smfig:phases} and Fig.~\ref{fig:real-space samples} of the main text.

\begin{figure*}[t]
\begin{center}
\includegraphics[width=0.8\textwidth]{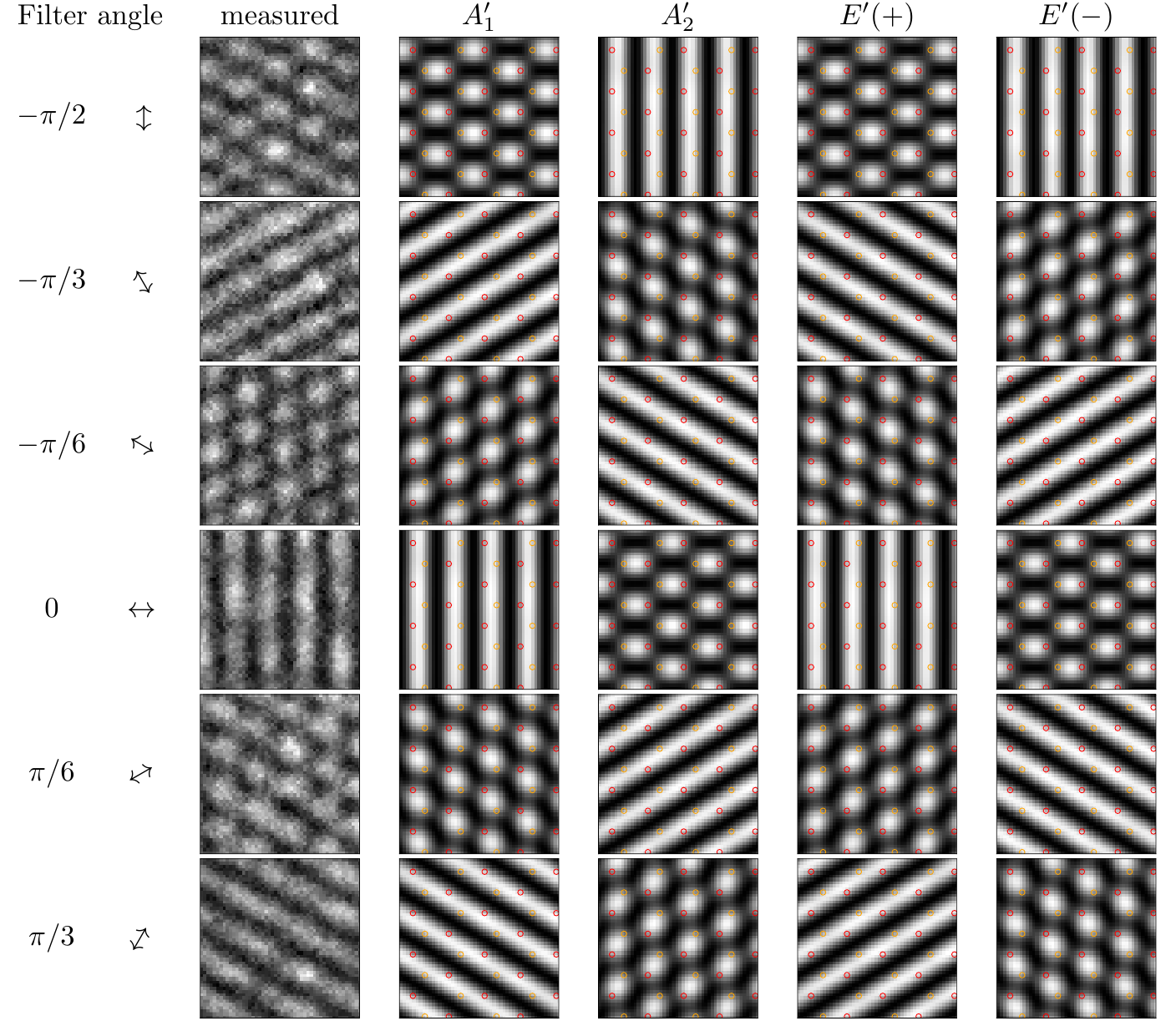}
\end{center}
\caption{Comparison of measured real space patterns (left column) with the
	theoretically predicted patterns for various eigenstate choices (the
	rest of the columns), for different output emission polarization filter
	angles. Here the angle zero corresponds to the horizontal polarization
	filter in Fig.~\ref{smfig:wholearrays} and Fig.~\ref{smfig:phases}, and
	in Fig.~\ref{fig:real-space samples} of the main manuscript. The theoretical predictions are for
	the singlets \TheSinglet{} and $A_2'$ (with uncorrelated phase between the
	$\Kp$ and $\Kp'$ contributions in both cases), and for the doublet state
	$E'$, for two different choices of the superposition phase between the
	doublet states (sum and difference, respectively, of the two doublet
	states depicted in Fig.~\ref{fig:eigenmodes theory}(a) in the main text).}
\label{smfig:comparison}
\end{figure*}

\subsection{Phase dependence of the real-space patterns}\label{smfig:phase}

Fig.~\ref{smfig:phases} shows how the simulated interference patterns evolve
when the relative phase between the $\Kp$- and $\Kp$'-point lasing contributions
vary. The situation where the relative phase is random produces distinctly
different interference patterns. These results demonstrate that the interference
patterns can serve as accurate probe of not only the lasing modes involved but
also of their relative phases.

\begin{figure*}[t]
\begin{center}
\includegraphics[width=0.9\textwidth]{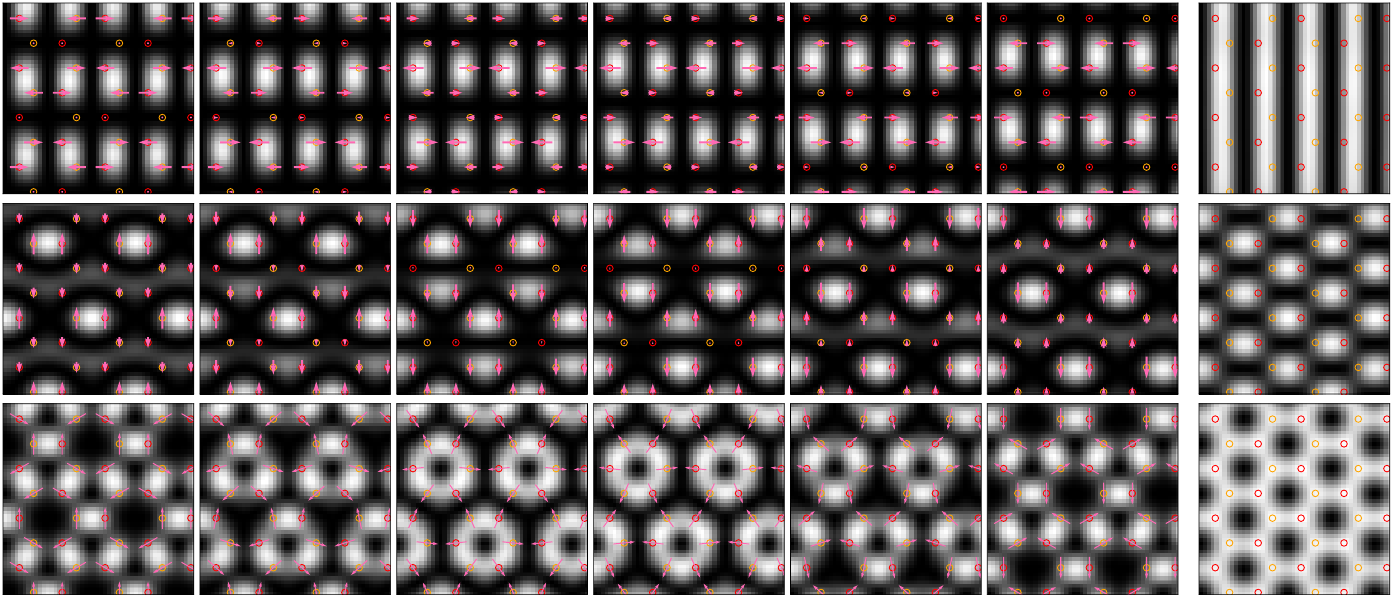}
\end{center}
\caption{Dependence of the real-space patterns on the relative phase of the $K$
	and $K'$ point realisations of the \TheSinglet{} mode. From top to bottom:
	horizontal filter, vertical filter, unfiltered. The sequences on the
	left depict the evolution of the real space pattern if the relative
	phase is shifted up to $\pi/3$. The patterns on the right correspond to
	the averaged intensity if the relative phase is totally random (or if
	only one of the $K$ and $K'$-modes contribute).}
\label{smfig:phases}
\end{figure*}

\section{$T$-matrix simulations\label{sm:tmatrix}}

\begin{figure*}
	\begin{center}
		\includegraphics[width=0.95\textwidth]{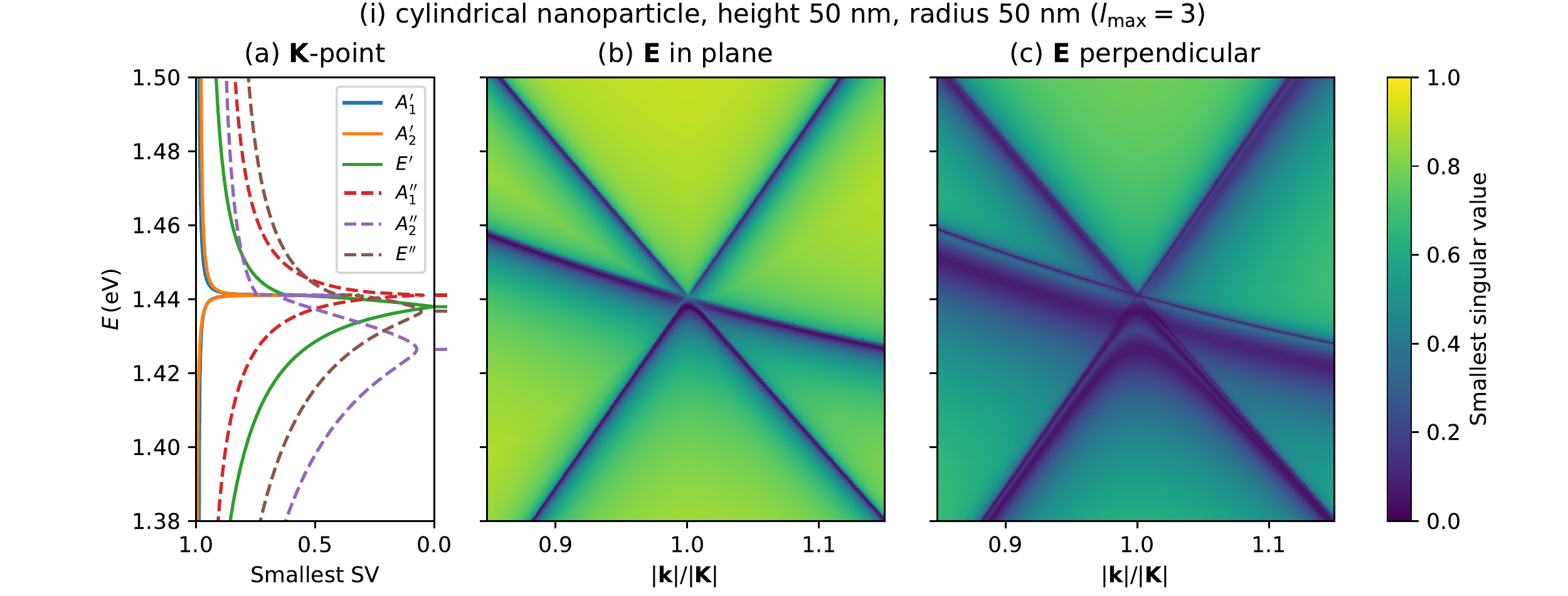}\\
		\vspace{6pt}
		\includegraphics[width=0.95\textwidth]{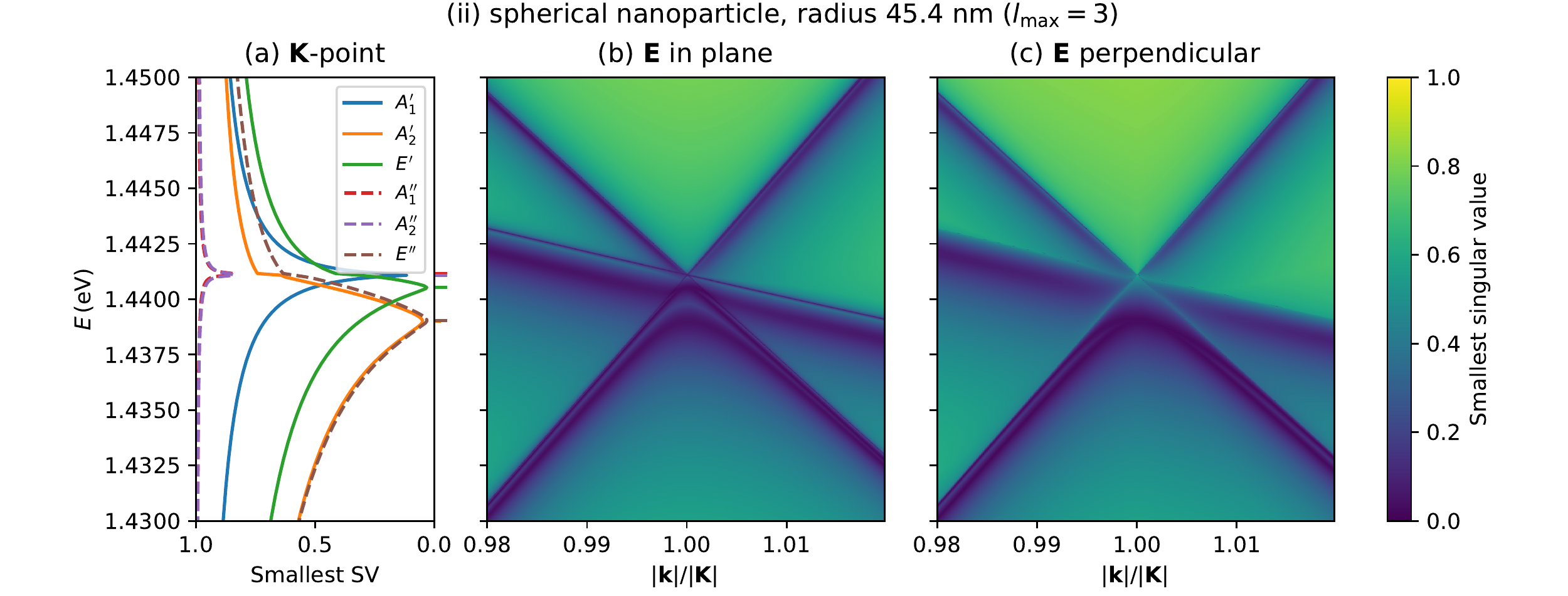}\\
	\end{center}
	\caption{{Band structure of infinite arrays around the $\Kp$-point obtained
using the $T$-matrix approach, with (i) $T$-matrix for a cylindrical
nanoparticle (height 50 nm, radius 50 nm) computed with BEM, and (ii)
$T$-matrix for a spherical nanoparticle (radius 45.4 nm) calculated
using Lorenz-Mie theory. The lowest singular value \changed{(SV)} of (\ref{eq:M matrix definition})
as a function of $(\omega,\protect\vect k)$ is shown 
	(a) exactly at the $\Kp$-point for each irrep separately,
	(b) for $\vect{E}$-in-plane modes, and 
(c) for $\vect{H}$-in-plane modes.}
\label{smfig:dispersions}
}

\end{figure*}

\begin{figure}
	\begin{center}
		\includegraphics[width=0.95\columnwidth]{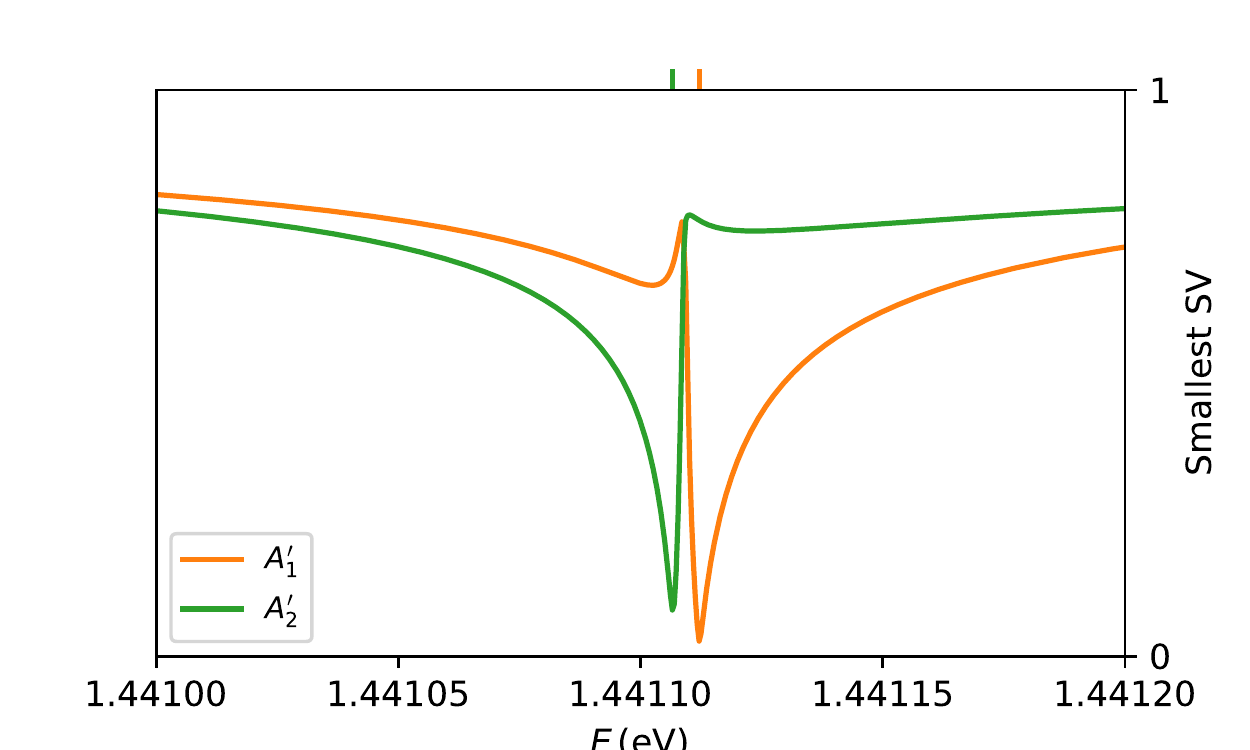}
	\end{center}
	\caption{\diffnote{NEW FIGURE.} \changed{The lowest singular values of (\ref{eq:M matrix definition}) exactly at the $\Kp$-point, 
	in the \TheSinglet{} and $A_2'$ subspaces for the cylindrical nanoparticle. The data are the same
	as in Figs. \ref{fig:eigenmodes theory}(e) and \ref{smfig:dispersions}(i)(a), but plotted on a scale
	that enables to distinguish between the two curves.}\label{smfig:dispersion detail}
	}
\end{figure}

In order to get more detailed insight into the mode structure of the
lattice around the lasing $\Kp$-point – most importantly, how much
do the mode frequencies at the $\Kp$-points differ from the empty
lattice model – we performed multiple-scattering $T$-matrix simulations
\cite{mackowski_analysis_1991} for an infinite lattice based on our
systems' geometry. {We give a brief overview of this method in the
subsections \ref{sub:The-multiple-scattering-problem}, \ref{sub:Periodic-systems}
below. The top advantage of the multiple-scattering $T$-matrix approach
is its computational efficiency for large finite systems of nanoparticles.
In the lattice mode analysis in this work, however, we use it here
for another reason, specifically the relative ease of describing symmetries
\cite{schulz_point-group_1999}.}

%\begin{longchange}
Fig. \ref{smfig:dispersions}(i) shows the dispersions around the
$\Kp$-point for the cylindrical nanoparticles used in our experiment.
The $T$-matrix of a single cylindrical nanoparticle was computed
using the scuff-tmatrix application from the SCUFF-EM suite~\cite{SCUFF2,reid_efficient_2015}
and the system was solved up to the $l_{\mathrm{max}}=3$ (octupolar)
degree of electric and magnetic spherical multipole. For comparison,
Fig. \ref{smfig:dispersions}(ii) shows the dispersions for a system
where the cylindrical nanoparticles were replaced with spherical ones
with radius of $45.4\,\mathrm{nm}$, whose $T$-matrix was calculated
semi-analytically using the Lorenz-Mie theory. In both cases, we used
gold with interpolated tabulated values of refraction index \cite{johnson_optical_1972}
for the nanoparticles and constant refraction index of 1.52 for the
background medium. In both cases, the diffracted orders do split into
separate bands according to the $\Kp$-point irreducible representations
(cf. section \ref{sm:symmetries}), but
the splitting is weak – not exceeding $2\,\mathrm{meV}$
for the spherical and 15 meV (3.2 meV for the $\vect{E}$-in-plane modes)
for the cylindrical nanoparticles. \changed{The splitting between \TheSinglet{}
and $A_2'$ is very small; Fig.~\ref{smfig:dispersion detail} shows
a detail from Fig.~\ref{fig:eigenmodes theory}(e) on a scale that
enables to distinguish them.}

\subsection{The multiple-scattering problem\label{sub:The-multiple-scattering-problem}}

In the $T$-matrix approach, scattering properties of single nanoparticles 
\changed{in a homogeneous medium} 
are first computed in terms of vector sperical wavefunctions (VSWFs)—the
field incident onto the $n$-th nanoparticle from external sources
can be expanded as 
\begin{equation}
\vect E_{n}^{\mathrm{inc}}(\vect
r)=\sum_{l=1}^{\infty}\sum_{m=-l}^{+l}\sum_{t=\mathrm{E},\mathrm{M}}\coeffrip
nlmt\svwfr lmt\left(\vect r_{n}\right)\label{eq:E_inc} \end{equation}
where $\vect r_{n}=\vect r-\vect R_{n}$, $\vect R_{n}$ being the
position of the centre of $n$-th nanoparticle and $\svwfr lmt$ are
the regular VSWFs which can be expressed in terms of regular spherical
Bessel functions of $j_{k}\left(\left|\vect r_{n}\right|\right)$
and spherical harmonics $\ush km\left(\hat{\vect r}_{n}\right)$;
the expressions\changed{, together with a proof that the SVWFs span all 
the solutions of vector Helmholtz equation around the particle, justifying the expansion,} can be found e.g. in
	\cite[chapter 7]{kristensson_scattering_2016-2} (care must be taken
because of varying normalisation and phase conventions). On the other
hand, the field scattered by the particle can be (outside the particle's
circumscribing sphere) expanded in terms of singular VSWFs $\svwfs lmt$
which differ from the regular ones by regular spherical Bessel functions
being replaced with spherical Hankel functions $h_{k}^{(1)}\left(\left|\vect r_{n}\right|\right)$,
\begin{equation}
\vect E_{n}^{\mathrm{scat}}\left(\vect r\right)=\sum_{l,m,t}\coeffsip nlmt\svwfs lmt\left(\vect r_{n}\right).\label{eq:E_scat}
\end{equation}
The expansion coefficients $\coeffsip nlmt$, $t=\mathrm{E},\mathrm{M}$
are related to the electric and magnetic multipole polarization amplitudes
of the nanoparticle. 

At a given frequency, assuming the system is linear, the relation
between the expansion coefficients in the VSWF bases is given by the
so-called $T$-matrix, 
\begin{equation}
\coeffsip nlmt=\sum_{l',m',t'}T_{n}^{lmt;l'm't'}\coeffrip n{l'}{m'}{t'}.\label{eq:Tmatrix definition}
\end{equation}
The $T$-matrix is given by the shape and composition of the particle
and fully describes its scattering properties. In theory it is infinite-dimensional,
but in practice (at least for subwavelength nanoparticles) its elements
drop very quickly to negligible values with growing degree indices
$l,l'$, enabling to take into account only the elements up to some
finite degree, $l,l'\le l_{\mathrm{max}}$. The $T$-matrix can be
calculated numerically using various methods; here we used the scuff-tmatrix
tool from the SCUFF-EM suite \cite{SCUFF2,reid_efficient_2015}\MMNmark{, 
which implements the boundary element method (BEM)}.

The singular VSWFs originating at $\vect R_{n}$ can be then re-expanded
around another origin (nanoparticle location) $\vect R_{n'}$ in terms
of regular VSWFs,
\begin{equation}
\begin{split}\svwfs lmt\left(\vect r_{n}\right)=\sum_{l',m',t'}\transop^{l'm't';lmt}\left(\vect R_{n'}-\vect R_{n}\right)\svwfr{l'}{m'}{t'}\left(\vect r_{n'}\right),\\
\left|\vect r_{n'}\right|<\left|\vect R_{n'}-\vect R_{n}\right|.
\end{split}
\label{eq:translation op def}
\end{equation}
Analytical expressions for the translation operator $\transop^{lmt;l'm't'}\left(\vect R_{n'}-\vect R_{n}\right)$
can be found in \cite{xu_efficient_1998}.

If we write the field incident onto the $n$-th nanoparticle as the sum
of fields scattered from all the other nanoparticles and an external
field $\vect E_{0}$ \changed{(which we also expand around each nanoparticle,
$\vect E_{0}\left(\vect r\right)=\sum_{l,m,t}\coeffripext nlmt\svwfr lmt\left(\vect r_{n}\right)$)}, 
\[
\vect E_{n}^{\mathrm{inc}}\left(\vect r\right)=\vect E_{0}\left(\vect r\right)+\sum_{n'\ne n}\vect E_{n'}^{\mathrm{scat}}\left(\vect r\right)
\]
and use eqs. (\ref{eq:E_inc})–(\ref{eq:translation op def}), we
obtain a set of linear equations for the electromagnetic response
(multiple scattering) of the whole set of nanoparticles,
\begin{equation}
\begin{split}\coeffrip nlmt=\coeffripext nlmt+\sum_{n'\ne n}\sum_{l',m',t'}\transop^{lmt;l'm't'}\left(\vect R_{n}-\vect R_{n'}\right)\\
\times\sum_{l'',m'',t''}T_{n'}^{l'm't';l''m''t''}\coeffrip{n'}{l''}{m''}{t''}.
\end{split}
\label{eq:multiplescattering element-wise}
\end{equation}%
\removed{where $\coeffripext nlmt$ are the expansion coefficients of the external
field around the $n$-th particle, $\vect E_{0}\left(\vect r\right)=\sum_{l,m,t}\coeffripext nlmt\svwfr lmt\left(\vect r_{n}\right).$}
It is practical to get rid of the VSWF indices, rewriting (\ref{eq:multiplescattering element-wise})
in a per-particle matrix form
\begin{equation}
\coeffr_{n}=\coeffr_{\mathrm{ext}(n)}+\sum_{n'\ne n}S_{n,n'}T_{n'}p_{n'}\label{eq:multiple scattering per particle p}
\end{equation}
and to reformulate the problem using (\ref{eq:Tmatrix definition})
in terms of the $\coeffs$-coefficients which describe the multipole
excitations of the particles
\begin{equation}
\coeffs_{n}-T_{n}\sum_{n'\ne n}S_{n,n'}\coeffs_{n'}=T_{n}\coeffr_{\mathrm{ext}(n)}.\label{eq:multiple scattering per particle a}
\end{equation}
Knowing $T_{n},S_{n,n'},\coeffr_{\mathrm{ext}(n)}$, the nanoparticle
excitations $a_{n}$ can be solved by standard linear algebra methods.
The total scattered field anywhere outside the particles' circumscribing
spheres is then obtained by summing the contributions (\ref{eq:E_scat})
from all particles.

\subsection{Periodic systems and mode analysis\label{sub:Periodic-systems}}

In an infinite periodic array of nanoparticles, the excitations of
the nanoparticles take the quasiperiodic Bloch-wave form
\[
\coeffs_{i\nu}=e^{i\vect k\cdot\vect R_{i}}\coeffs_{\nu}
\]
(assuming the incident external field has the same periodicity, $\coeffr_{\mathrm{ext}(i\nu)}=e^{i\vect k\cdot\vect R_{i}}p_{\mathrm{ext}\left(\nu\right)}$)
where $\nu$ is the index of a particle inside one unit cell and
$\vect R_{i},\vect R_{i'}\in\Lambda$ are the lattice vectors corresponding
to the sites (labeled by multiindices $i,i'$) of a Bravais lattice
$\Lambda$. The multiple-scattering problem (\ref{eq:multiple scattering per particle a})
then takes the form

\[
\coeffs_{i\nu}-T_{\nu}\sum_{(i',\nu')\ne\left(i,\nu\right)}S_{i\nu,i'\nu'}e^{i\vect k\cdot\left(\vect R_{i'}-\vect R_{i}\right)}\coeffs_{i\nu'}=T_{\nu}\coeffr_{\mathrm{ext}(i\nu)}
\]
or, labeling $W_{\nu\nu'}=\sum_{i';(i',\nu')\ne\left(i,\nu\right)}S_{i\nu,i'\nu'}e^{i\vect k\cdot\left(\vect R_{i'}-\vect R_{i}\right)}=\sum_{i';(i',\nu')\ne\left(0,\nu\right)}S_{0\nu,i'\nu'}e^{i\vect k\cdot\vect R_{i'}}$
and using the quasiperiodicity,
\begin{equation}
\sum_{\nu'}\left(\delta_{\nu\nu'}\mathbb{I}-T_{\nu}W_{\nu\nu'}\right)\coeffs_{\nu'}=T_{\nu}\coeffr_{\mathrm{ext}(\nu)},\label{eq:multiple scattering per particle a periodic}
\end{equation}
which reduces the linear problem (\ref{eq:multiple scattering per particle a})
to interactions between particles inside single unit cell. A problematic
part is the evaluation of the translation operator lattice sums $W_{\nu\nu'}$;
this is performed using exponentially convergent Ewald-type representations
\cite{linton_lattice_2010}.

In an infinite periodic system, a nonlossy mode supports itself without
external driving, i.e. such mode is described by excitation coefficients
$a_{\nu}$ that satisfy eq. (\ref{eq:multiple scattering per particle a periodic})
with zero right-hand side. That can happen if the block matrix 
\begin{equation}
M\left(\omega,\vect k\right)=\left\{ \delta_{\nu\nu'}\mathbb{I}-T_{\nu}\left(\omega\right)W_{\nu\nu'}\left(\omega,\vect k\right)\right\} _{\nu\nu'}\label{eq:M matrix definition}
\end{equation}
 from the left hand side of (\ref{eq:multiple scattering per particle a periodic})
is singular (here we explicitely note the $\omega,\vect k$ depence).

For lossy nanoparticles, however, perfect propagating modes will not
exist and $M\left(\omega,\vect k\right)$ will never be perfectly
singular. Therefore in practice, we get the bands by scanning over
$\omega,\vect k$ to search for $M\left(\omega,\vect k\right)$ which
have an ''almost zero'' singular value.

\section{{Symmetries}}\label{sm:symmetries}

A general overview of utilizing group theory to find lattice modes
at high-symmetry points of the Brillouin zone can be found e.g. in
\cite[chapters 10–11]{dresselhaus_group_2008}; here we use the
same notation.

We analyse the symmetries of the system in the same VSWF representation
as used in the $T$-matrix formalism introduced above. We are interested
in the modes at the $\Kp$-point of the hexagonal lattice, which has
the $D_{3h}$ point symmetry.  The six irreducible representations
(irreps) of the $D_{3h}$ group are known and are available in the
literature in their explicit forms. In order to find and classify
the modes, we need to find a decomposition of the lattice mode representation
$\Gamma_{\mathrm{lat.mod.}}=\Gamma^{\mathrm{equiv.}}\otimes\Gamma_{\mathrm{vec.}}$
into the irreps of $D_{3h}$. The equivalence representation $\Gamma^{\mathrm{equiv.}}$
is the $E'$ representation as can be deduced from \cite[eq. (11.19)]{dresselhaus_group_2008},
eq. (11.19) and the character table for $D_{3h}$. $\Gamma_{\mathrm{vec.}}$
operates on a space spanned by the VSWFs around each nanoparticle
in the unit cell (the effects of point group operations on VSWFs are
described in \cite{schulz_point-group_1999}). This space can be then
decomposed into invariant subspaces of the $D_{3h}$ using the projectors
$\hat{P}_{ab}^{\left(\Gamma\right)}$ defined by \cite[eq. (4.28)]{dresselhaus_group_2008}.
This way, we obtain a symmetry adapted basis $\left\{ \vect b_{\Gamma,r,i}^{\mathrm{s.a.b.}}\right\} $
as linear combinations of VSWFs $\svwfs lm{p,t}$ around the constituting
nanoparticles (labeled $p$),
\[
\vect b_{\Gamma,r,i}^{\mathrm{s.a.b.}}=\sum_{l,m,p,t}U_{\Gamma,r,i}^{p,t,l,m}\svwfs lm{p,t},
\]
where $\Gamma$ stands for one of the six different irreps of $D_{3h}$,
$r$ labels the different realisations of the same irrep, and the
last index $i$ going from 1 to $d_{\Gamma}$ (the dimensionality
of $\Gamma$) labels the different partners of the same given irrep.
The number of how many times is each irrep contained in $\Gamma_{\mathrm{lat.mod.}}$
(i.e. the range of index $r$ for given $\Gamma$) depends on the
multipole degree cutoff $l_{\mathrm{max}}$. 

Each mode at the $\Kp$-point shall lie in the irreducible spaces
of only one of the six possible irreps and it can be shown via \cite[eq. (2.51)]{dresselhaus_group_2008}
that, at the $\Kp$-point, the matrix $M\left(\omega,\vect k\right)$
defined above takes a block-diagonal form in the symmetry-adapted
basis,
\[
M\left(\omega,\vect K\right)_{\Gamma,r,i;\Gamma',r',j}^{\mathrm{s.a.b.}}=\frac{\delta_{\Gamma\Gamma'}\delta_{ij}}{d_{\Gamma}}\sum_{q}M\left(\omega,\vect K\right)_{\Gamma,r,q;\Gamma',r',q}^{\mathrm{s.a.b.}}.
\]
 This enables us to decompose the matrix according to the irreps and
to solve the singular value problem in each irrep separately, as done
in Fig. \ref{smfig:dispersions}(a).

%\end{longchange}

%\bibliography{Refs,hexarray-theory}


\end{document}